\documentclass[notitlepage,a4paper,aps,prd,onecolumn,superscriptaddress,nofootinbib,longbibliography,20pt]{revtex4}
\usepackage{amsthm}
\usepackage{makeidx}
\usepackage{graphicx}
\usepackage{microtype}
\usepackage{amsmath}
\usepackage{amssymb}
\usepackage{subfigure}
\usepackage{hyperref}
\usepackage{url}
\usepackage{adjustbox}
\usepackage{xcolor}
\usepackage{color}
\usepackage{mathrsfs}
\usepackage{calrsfs}
\usepackage{amsfonts}
\usepackage{lipsum}
\usepackage{eufrak}
\usepackage{tabularx}
\usepackage{eucal}
\usepackage{float}
\usepackage{latexsym}
\usepackage{ragged2e}
\usepackage{epsfig}
\usepackage{textcomp}
\usepackage{multirow}
\usepackage{booktabs}
\usepackage{caption}
\usepackage{orcidlink}
\usepackage{hyperref} 
\usepackage{xcolor}
\usepackage[normalem]{ulem}

\DeclareCaptionJustification{justified}{\leftskip=0pt \rightskip=0pt \parfillskip=0pt plus 1fil}
\usepackage{calligra}
\DeclareMathAlphabet{\mathcalligra}{T1}{calligra}{m}{n}
\DeclareFontShape{T1}{calligra}{m}{n}{<->s*[2.2]callig15}{}

\definecolor{vividviolet}{rgb}{0.62, 0.0, 1.0}
\definecolor{amaranth}{rgb}{0.9, 0.17, 0.31}
\definecolor{palatinateblue}{rgb}{0.15, 0.23, 0.89}
\definecolor{brightpink}{rgb}{1.0, 0.0, 0.5}
\definecolor{cornflowerblue}{rgb}{0.39, 0.58, 0.93}
\definecolor{deepcarminepink}{rgb}{0.94, 0.19, 0.22}
\definecolor{radicalred}{rgb}{1.0, 0.21, 0.37}

\hypersetup{ linktoc=all,
	colorlinks, linkcolor={palatinateblue},
	citecolor={brightpink}, urlcolor={amaranth}
}

\graphicspath{{Images/}}

\def\sideremark#1{\ifvmode\leavevmode\fi\vadjust{\vbox to0pt{\vss
			\hbox to 0pt{\hskip\hsize\hskip1em
				\vbox{\hsize1.3cm\tiny\raggedright\pretolerance10000
					\noindent #1\hfill}\hss}\vbox to8pt{\vfil}\vss}}}%
\def\beq{\begin{equation}}
\def\eeq{\end{equation}}
\newcommand{\be}{\begin{equation}}
\newcommand{\ee}{\end{equation}}
\newcommand{\ba}{\begin{eqnarray}}
\newcommand{\ea}{\end{eqnarray}}

\begin{document}

\title{Unified dynamical system formulations for $f(R,\phi,X)$ gravity with applications to nonminimal derivative coupling and $R^2$-Higgs inflation}

\author{Saikat Chakraborty\orcidlink{0000-0002-5472-304X}}
	\email{saikat.chakraborty@nwu.ac.za}
	\thanks{Corresponding author}
	\affiliation{Institute of Research and Development, Duy Tan University, Da Nang 550000, Vietnam}
    \affiliation{Faculty of Natural Sciences, Duy Tan University, Da Nang 550000, Vietnam}
	\affiliation{Center for Space Research, North-West University, Potchefstroom 2520, South Africa}

\author{Alberto Saa\orcidlink{0000-0003-1520-4076}} 
\email{asaa@ime.unicamp.br}
\affiliation{Department of Applied Mathematics, University of Campinas, 13083-859,
Campinas, SP, Brazil}

\author{Sergio E. Jor\'as\orcidlink{0000-0002-4627-2111}}
\email{joras@if.ufrj.br}
\affiliation{Instituto de F\'\i sica, Universidade Federal do Rio de Janeiro, 21941-972, RJ, Brazil}

\begin{abstract}
Two different dynamical system formulations are presented that can be implemented to analyze a rather large class of modified gravity theories within the generic $f(R,\phi,X)$ family. As illustrative examples, the first and the second formulation is applied to study the phase space of a toy model of the Non-Minimal Derivative Coupling (NMDC) without a potential, and the mixed $R^2$-Higgs inflation model, respectively. The first dynamical system formulation applied to the toy NMDC model, although able to identify several invariant submanifolds, fails to fully investigate the fixed point structure, as all the fixed points turn out to be non-hyperbolic. We, however, discover an interesting feature that the qualitative dynamics are independent of the coupling strength between the Ricci scalar and the scalar field derivative for the particular toy model under consideration. The second dynamical system formulation applied to the mixed $R^2$-Higgs inflation model performs much better, being able to correctly reduce to the individual phase spaces of the $R^2$ and Higgs inflation separately in special cases, as well as correctly delivering the expected invariant submanifolds and fixed points. For the mixed $R^2$-Higgs case, illustrative phase portraits are provided for a somewhat better visual understanding of the dynamics. 
\end{abstract}

\maketitle

\section{Introduction}

The conception of the inflationary paradigm of the early universe and the discovery of the acceleration of the late-time universe have acted as two catalysts for intense research into extensions of General Relativity (GR). Among the diverse landscape of modified gravity theories, two classes have garnered significant attention: $f(R)$ gravity, which generalizes the Einstein-Hilbert action by introducing an arbitrary function of the Ricci scalar, and scalar-tensor theories, which introduce non-minimal couplings between the Ricci scalar and a cosmological scalar field $\phi$ (e.g., the generalized Brans-Dicke theory). Later, these two theories were grouped under the general $f(R,\phi)$ class of gravity theories, which have been extensively studied. With the introduction of notions like noncanonical kinetic terms \cite{Armendariz-Picon:1999hyi,Armendariz-Picon:2000ulo} and derivative coupling of the field to the geometry \cite{Capozziello:1999uwa,Granda:2009fh}, an even bigger generalized class of gravity theories can be conceived; the $f(R,\phi,X)$ class of theories, where the Lagrangian is a general function of the Ricci scalar $R$, a scalar field $\phi$, and its kinetic term $X = \frac{1}{2} \partial_\mu \phi \partial^\mu \phi$. This generic family encapsulates a vast array of phenomenologically viable models, including $f(R)$ gravity \cite{Amendola:2006we,Carloni:2007br}, non-minimally coupled scalar fields \cite{Faraoni:2000nt,Faraoni:2004dn,Faraoni:2005ie,Faraoni:2005vk,Gunzig:2000kk}, models with tachyonic condensation \cite{Mazumdar:2001mm,Gibbons:2002md,Choudhury:2002xu,Feinstein:2002aj,Padmanabhan:2002cp} etc, offering a broad arena for exploring deviations from GR. However, not all such theories may be written in simpler forms like $R+f(\phi,X)$ or $f(R,\phi)+X$. Different specific forms of Lagrangians belonging to the $f(R,\phi,X)$ class have been studied in \cite{Shamir:2020oyu,Bahamonde:2015shn,Keskin:2018awq,Myrzakulov:2015qaa,Cecchini:2024xoq}

It is important to recognize that the $f(R,\phi, X)$ class of theories is most naturally interpreted as a low-energy effective theory of gravity rather than a fundamental description itself. Such an action encompasses higher-curvature terms, nonminimal couplings, and noncanonical kinetic structures that generically arise once heavy degrees of freedom are integrated out or quantum corrections are taken into account in curved spacetime. For a comprehensive review of effective field theory (EFT),
see \cite{Buchbinder:1992rb}. An important cosmological application is the Born-Infeld action \cite{Tseytlin:1999dj}, which has inspired many of the models recently considered in \cite{Shamir:2020oyu,Bahamonde:2015shn,Keskin:2018awq,Myrzakulov:2015qaa,Cecchini:2024xoq}. From an EFT standpoint, the generality of the $f(R,\phi, X)$ class of theories is a strength: it provides a unifying language in which phenomenologically relevant models, such as $f(R)$ gravity, nonminimal derivative couplings, non-standard kinetic terms, and $R^2$-Higgs inflation, can be treated on the same footing and analyzed systematically at the level of background cosmology. At the same time, this very generality also exposes its limitations as a fundamental theory. Higher derivatives often indicate the presence of additional propagating degrees of freedom, which can lead to concerns regarding ghosts or a loss of predictability, unless it is understood that the theory is only valid below a certain cutoff scale.

One of our motivations for a generic dynamical analysis of the $f(R,\phi, X)$ class of theories comes largely from the works of Hwang and Noh, who had earlier performed some generic cosmological perturbation analysis on this broad class of theories \cite{Hwang:2002fp, Hwang:2005hb}. However, a significant challenge in studying such a generic class of theories is precisely the absence of a suitable set of dynamical variables. The dynamical systems analysis we have developed here should be viewed as a probe of the cosmological phase space within the EFT regime with appropriate dynamical variables, aiming to identify robust attractors and invariant submanifolds that are insensitive to ultraviolet completion, while remaining agnostic about the ultimate high-energy origin of the theory.

Given the nonlinearity of the field equations in relativistic gravity theories, finding realistic, exact analytical solutions for cosmological evolution is next to impossible. Consequently, dynamical system analysis has emerged as an indispensable mathematical tool for qualitative cosmology. By transforming the cosmological evolution equations into a system of autonomous ordinary differential equations, one can map the entire phase space of the theory. This approach allows for the identification of fixed points, representing critical cosmological epochs such as inflation, radiation domination, matter domination, or a de Sitter acceleration. The utility of this method has been demonstrated extensively in the literature, providing robust constraints on the viability of various dark energy models \cite{Bahamonde:2017ize}. In particular, the formulation has seen extensive and very successful application in the context of $f(R)$ gravity \cite{Amendola:2006we,Carloni:2007br,Chakraborty:2021hmx,MacDevette:2022hts,Chakraborty:2018bij,Chakraborty:2018qew}

Despite the wealth of studies focusing on isolated subclasses of modified gravity theories, some of which also consider a scalar field with noncanonical kinetic terms \cite{Panpanich:2021lsd,Moretti:2025dof}, a unified dynamical system formulation for the generic $f(R,\phi,X)$ action with universal applicability is notably absent from the current literature as of yet. Previous works have largely focused on $f(R)$ modifications and the dynamics of different scalar field models. The Ref.\cite{Bahamonde:2017ize} provides a very nice review. There is, therefore, a clear need for a master framework that can treat $f(R,\phi,X)$ gravity in its full generality. Such a formulation would not only recover standard results for specific limits---such as quintessence or pure $f(R)$ gravity---but also unveil novel phase space features arising from the interplay between the scalar kinetic term and the non-linear curvature sector.

In this paper, we propose two unified dynamical system frameworks for homogeneous and isotropic FLRW cosmology within the $f(R,\phi,X)$ class of gravity theories. By carefully selecting a set of dimensionless variables, we construct an autonomous system that governs the background evolution for this broad family of models. As we will see, although the first formulation employs dynamical variables that are exactly parallel to the ones used in $f(R)$ gravity dynamical system formulation, its success is subject to two conditions. We carefully identify both of them: an invertibility condition and a nondegeneracy one. These limitations motivate us to propose a second, alternative formulation, which, although having its own nondegeneracy condition, is free from any invertibility issue. Both formulations are built upon the assumptions of $f_R\neq0$ and $f_X\neq0$ identically. The dynamical system in the first formulation becomes singular \emph{locally} if any of $\{R,f,f_R,f_X\}$ vanish at a given point, while that in the second formulation becomes singular \emph{locally} if any of $\{R,,f_R,f_X\}$ vanish momentarily. A dynamical system becoming singular locally implies the corresponding physical situations will not be captured by the particular formulation. We have been explicit about these applicability limitations in the text. However, even with these limitations, the formulations do allow us to investigate some physically interesting models. The absence of an invertibility condition and the regularity at $f=0$ represent two distinct advantages of the second formulation as compared to the first formulation.

Our principal motivation behind this work is that either one or both of these unified approaches should allow us to study the phase space of all possible cosmological models that are of the form $f(R,\phi,X)$. Hence, one will not need to spend the effort of devising a dynamical system formulation for any individual subclass of such models under consideration. 

To demonstrate the applicability of our formulations in studying practical cosmological models, we employ the two formulations to study the phase space of two different models. The first formulation is applied to study the phase space, as much as possible, of a toy model with nonminimal derivative coupling, which is a specific case of the model appearing in \cite{Capozziello:1999uwa}. The second formulation is applied to study the phase space of mixed $R^2$-Higgs inflation \cite{He:2018gyf}.

The paper is organized as follows. In Section \ref{Sec: cosmology} we outline FLRW cosmology in $f(R,\phi,X)$ gravity. In Section \ref{Sec: DS_1} and \ref{Sec: DS_2}, we present the two dynamical system formulations, respectively. In Section \ref{Sec: NMDC}, the first formulation is employed to study a toy NMDC model as much as possible. In Section \ref{Sec: SH}, the second formulation is employed to study the mixed $R^2$-Higgs inflation. Our findings are summarized in Section \ref{summary}. Finally, Appendix \ref{app:f(R)} is dedicated to showing explicitly that the first dynamical system formulation correctly reduces to the well-known $f(R)$ dynamical system for a frozen scalar field, and the Appendix \ref{app:SH} is dedicated to explaining an interesting feature of the mixed $R^2$-Higgs model.

\section{Cosmological field equations}\label{Sec: cosmology}

We are interested in the  homogeneous and isotropic cosmological solutions of a  generic $f(R,\phi,X)$ modified theory of gravity  
of the type
\begin{equation}\label{action}
S = \int d^4 x \sqrt{-g} \left(\frac{1}{2\kappa^2}f(R,\phi,X) + \mathcal{L}_m \right),
\end{equation}
where $X = \frac{1}{2}\partial_a\phi\partial^a\phi$\footnote{While working with the generalized gravity Lagrangian of the form $f(R,\phi,X)$, some authors prefer to use the convention $X=-\frac{1}{2}\partial_\alpha\phi\partial^{\alpha}\phi$. See, for example, the works \cite{Bahamonde:2015shn,Keskin:2018awq,Shamir:2020oyu}. To avoid getting confused with conventions, it is important to set a convention straight from the very beginning. We follow closely the convention of Hwang-Noh \cite{Hwang:2002fp,Hwang:2005hb}. Our equations can be tallied with those arising in the Hwang-Noh papers. However, we believe on the right-hand side of \cite[Eq.(8)]{Hwang:2002fp}, there should be $8\pi G\mu^{(m)}$ and $8\pi G p^{(m)}$ instead of just $\mu^{(m)}$ and $p^{(m)}$ as appears there. We accordingly amend it here. We also believe there is a $\mu^{(m)},\,p^{(m)}$ missing on the right-hand side of the definitions of the effective fluid in Eq.(77) of \cite{Hwang:2005hb}. To verify our claims, the reader can tally the Hwang-Noh equations with those arising in the other works \cite{Bahamonde:2015shn,Keskin:2018awq,Shamir:2020oyu}.}. The equations of motion for the FLRW metric
\begin{equation}
\label{metric}
ds^2 = -dt^2 + a^2(t)\left(\frac{dr^2}{1-k r^2} + r^2 d \Omega^2 \right),
\end{equation}
with $k=0,\pm 1,$ are  (see Eqs. (7)-(9) of \cite{Hwang:2002fp})
\begin{subequations}\label{field_eqs}
\begin{eqnarray}
&& 3f_R\left(H^2 + \frac{k}{a^2}\right) = \kappa^2\rho - \frac{1}{2}(f-f_RR)    - 3H\dot f_R + X f_X \,,\label{constr0}
\\
&& -2f_R\left(\dot H  - \frac{k}{a^2}\right) = \kappa^2(\rho + P) + \ddot f_R   - H\dot f_R + X f_X \,,\label{dyn0}
\\
&& \frac{d}{dt} \left(\dot\phi f_X  \right) = (f_X + 2Xf_{XX})\ddot{\phi} + f_{XR}\dot{\phi}\dot{R} - 2Xf_{X\phi} = - 3H\dot\phi f_X  -f_\phi \,,\label{KG0}
\end{eqnarray}
\end{subequations}
where $\kappa^2=8\pi G$, $\rho$ and $P$ stand for, respectively, the energy density and the pressure of the matter content,
which is
assumed to be a perfect fluid, 
  $H=\frac{\dot a}{a}$, with the dot denoting derivative with respect to $t$ and the subscript denoting
  a partial derivative. Also, for the metric (\ref{metric}), we have 
\begin{equation}\label{Rdef}
R = 6\dot H + 12 H^2 + \frac{6k}{a^2}.
\end{equation}
There is an $\dot{f}_R$ term in (\ref{constr0}) and (\ref{dyn0}). We can expand this term explicitly as
\begin{equation}
\label{dotf_R}
\dot f_R = f_{RR}\dot R + \left( f_{R\phi} - f_{RX} \ddot \phi\right)\dot \phi.
\end{equation}
One can use the Klein-Gordon  equation (\ref{KG0}) to express $\ddot \phi$ in terms of $H,\,R,\,\dot{R},\,\phi,\,\dot{\phi}$. Therefore, we stay finally with a function $\dot f_R = \dot f_R(R,\dot R,\phi,\dot \phi)$. 

There is an $\ddot{f}_R$ term in Eq.\eqref{dyn0}. One can take another time derivative of the equation \eqref{dotf_R} and obtain $\ddot f_R = \ddot f_R(R,\dot R,\ddot R,\phi,\dot \phi)$. Therefore, Eq.\eqref{dyn0} will contain a $\ddot{R}$ term.

Taking into account (\ref{Rdef}), we see that 
in terms of the primitive variables $(a,H,\phi)$, the system (\ref{constr0})-(\ref{KG0})
is six-dimensional for general $f(R,\phi,X)$, since the energy constraint can be used to eliminate the variables $\rho$ and $p$ (given a hydrodynamic equation of state $p=p(\rho)$), and Eqs. (\ref{dyn0}) and (\ref{KG0}) are, respectively, equations for the third derivative of $H$ and for the second derivatives of $\phi$. 
Take into account (\ref{Rdef}), we see that the natural phase space variables for this system are $(a,H,R,\dot R,\phi,\dot\phi)$. However, following the more modern approach to dynamical system formulation of modified gravity (see. e.g., \cite{Amendola:2006we,Carloni:2007br}), we will rather try to design Hubble-normalized dimensionless dynamical variables in the next two sections.

\section{Dynamical system of $f(R,\phi,X)$ gravity: 1st formulation}\label{Sec: DS_1}
 
In this section, following the approach of \cite{Amendola:2006we,Carloni:2007br}, we design Hubble normalized dimensionless dynamical variables. Solutions like Minkowski or Einstein static solutions, for which $H=0$, cannot appear as finite fixed points in our analysis. For $H\ne 0$, it is more convenient to introduce the new evolution parameter $N = \log a(t)$ such that $\frac{d}{dN} = \frac{1}{H}\frac{d}{dt}$, with $'$ denoting the derivative  $\frac{d}{dN}$. 

Eq.\eqref{constr0} strongly suggests the introduction of the new dimensionless variables
\begin{eqnarray}
\label{variables}
\Omega  = \frac{\kappa^2\rho}{3f_RH^2}, \quad  
x = \frac{f}{6f_R H^2} ,\quad
y = \frac{R}{6H^2}, \quad 
z = \frac{{f_R}'}{f_R} = \frac{\dot{f_R}}{H f_R}, \quad \nonumber \\
K = \frac{k}{a^2H^2}, \quad
q = \kappa\phi, \quad
p = \kappa\phi' = \kappa\frac{\dot{\phi}}{H} .
\end{eqnarray}
These variables are the direct extension for our case of the variables introduced, for instance, in \cite{Amendola:2006we} or \cite{Carloni:2007br}. One can note that, $\Omega,\,x,\,y,\,z,\,K$ are exactly the same dynamical variables (modulo, perhaps, only the sign) as what appears in case of $f(R)$ gravity. The dynamical variable $q,\,p$ has recently been explored in \cite{Jarv:2021qpp} where the authors claim to find them more useful than the traditionally used set $(\phi,\dot{\phi})$ \citep{Amendola:1990nn} to investigate the phase space structure of nonminimal inflation.
 
In terms of the variables (\ref{variables}), the cosmological field equations can be equivalently expressed in terms of the 7-dimensional system
\begin{subequations}\label{DS}
\begin{eqnarray}
x' &=& x\left(g_1-z - 2(y-2-K) \right) \,,\\
y' &=& y\left(g_2 - 2(y-2-K) \right)\,, \\
z' &=& 1 - 3x + y + K - 3w\Omega - z(y + z - K)\,, \\
K' &=& -2K (y-1-K)\,, \\
\Omega' &=& \Omega(1 - 3w - 2y - z + 2K)\,, \\
q' &=& {p}\,, \\ 
p' &=& -\left(\frac{f_{XR}}{f_X}R' + \frac{f_{XX}}{f_X}X'\right)p - \frac{f_{X\phi}}{f_X}p^2 -\left(y+1-K\right) p - \frac{6{\kappa} f_\phi}{f_X R}y\,.
\end{eqnarray}
\end{subequations}
along with the Friedmann constraint
\begin{equation}
    z = -1 + \Omega - K - x + y - \frac{p^2 f_X}{6\kappa^2 f_R}.\label{const}
\end{equation}
In the above, we have defined
\begin{subequations}\label{g}
\begin{eqnarray}
&& g_2=\frac{R'}{R}\,,
\\
&& g_1 = \frac{f'}{f} = \frac{f_R}{f}R' + \frac{f_\phi}{f}\phi' + \frac{f_X}{f}X' = \frac{R f_R}{f}g_2 + \frac{f_\phi}{f}p + \frac{f_X}{f}X' \,.
\end{eqnarray}
\end{subequations}
The $p'$-equation above can also be conveniently expressed as
\begin{equation}\label{fxp'}
    (p f_X)' = -\left(y +1 -K \right)p f_X - \frac{6{\kappa}f_\phi}{R}y. 
\end{equation}

It is to be mentioned that in arriving at the system \eqref{DS}, we have assumed that $f_R\neq0$ and $f_X\neq0$ identically. $f_R\neq0$ is a justified assumption within the context of $f(R,\phi,X)$ theory; because otherwise there would not be any propagating graviton. $f_X=0$ is still possible in a scalar-tensor gravity theory; the prime example being $f(R)$ gravity. Such cases lie outside our treatment. However, notice that the irregularity of the dynamical system \eqref{DS} at $f_X=0$ creeps in through the $p'$-equation. If the field is assumed to be frozen into a potential minimum, then $q'=p=0$ identically. In such cases, neither $q$ nor $p$ is a dynamical variable anymore, and one need not worry about the $p'$-equation. In that case, the scalar field potential value effectively acts as a cosmological constant in the action. With this realization, it is possible to reduce the dynamical system \eqref{DS} to the well-known $f(R)$ dynamical system (see Appendix \ref{app:f(R)}), which serves as a nice consistency check.

Also, it is to be mentioned that, even for such a truly scalar tensor gravity theory, the dynamical system \eqref{DS} can still become singular \emph{locally} if $R$ or $f$ or $f_R$ or $f_X$ vanish \emph{locally}; such physical situations will not be captured by this particular dynamical system formulation.

Using the constraint \eqref{const}, we can eliminate one of the dynamical variables and obtain a reduced dynamical system. Eliminating $z$ we obtain
\begin{subequations}\label{DS_reduced}
\begin{eqnarray}
x' &=& x\left[g_1 - 1 + x - \Omega - 3(y-K-2) + \frac{p^2 f_X}{6\kappa^2 f_R}\right]\label{x'} \,,\\
y' &=& y\left[g_2 - 2(y-2-K) \right]\label{y'}\,, \\
\Omega' &=& \Omega\left[-4 - 3w - \Omega + x + \frac{p^2 f_X}{6\kappa^2 f_R} - 3(y-K-2)\right] \label{Omega'}\,\\
K' &=& -2K (y-1-K)\,, \label{K'} \\
q' &=& {p} \,,\label{q'}  \\ 
p' &=& -\left(\frac{f_{XR}}{f_X}R' + \frac{f_{XX}}{f_X}X'\right)p - \frac{f_{X\phi}}{f_X}p^2 -\left(y+1-K\right) p - \frac{6{\kappa} f_\phi}{f_X R}y\,.\label{p'} 
\end{eqnarray}    
\end{subequations}

The system \eqref{DS_reduced} is still not autonomous. To have an autonomous system, one needs to be able to write  all the terms appearing in the equations of motion
 as functions solely of
  the variables (\ref{variables}). 
Notice, first, that we have
\begin{eqnarray}
\label{X} 
X=-\frac{1}{2}\dot\phi^2 =  -\frac{Rp^2}{12 \kappa^2y},
\end{eqnarray}
and
\begin{equation}
\label{R}
\frac{Rf_R}{f} = \frac{y}{x}.
\end{equation}
Since $f = f(R,\phi,X)$ with $\phi = q$, 
we can, in principle, solve the two equations \eqref{X} and \eqref{R} to obtain $R=R(x,y,q,p)$ and $X=X(x,y,q,p)$. This allows us to write all the terms involving
$f$ and its partial derivatives of all orders appearing in Eq.\eqref{const} and Eq.\eqref{DS_reduced} as functions of the dynamical variables $x,\,y,\,q,\,p$ (\ref{variables}). 

We are now left with the terms $R'$ and $X'$ appearing in the functions $g_1$ and $g_2$ in (\ref{g}). If we can express $R',\,X'$ in terms of the dynamical variables, we close the system. To this goal, we proceed as follows. From the definition of the dynamical variable $z$, and the constraint equation \eqref{const}, one can write
\begin{equation}\label{z-decomp}
z = \frac{{f_R}'}{f_R} = \frac{f_{RR}}{f_R}R' + \frac{f_{R\phi}}{f_R}\frac{p}{\kappa} + \frac{f_{RX}}{f_R}X' = -1 + \Omega - K - x + y - \frac{p^2 f_X}{6\kappa^2 f_R}\,.
\end{equation}
On the other hand, multiplying both sides of the expression in Eq.\eqref{X} by $f_X^2$ and taking its derivative, a straightforward calculation gives 
\begin{equation}\label{find_R'}
Xf_X\left(2f_{XR} - \frac{f_X}{R}\right)R' + 2Xf_Xf_{X\phi}\frac{p}{\kappa} + (f_X^2 + 2Xf_Xf_{XX})X' = \left(\frac{2Xf_X}{p}\right)(p f_X)' - \left(\frac{Xf_X^2}{y}\right)y'\,.
\end{equation}
In writing the above equations, $\phi'=\frac{p}{\kappa}$ is used. Using Eqs.\eqref{fxp'} and \eqref{y'}, the two equations \eqref{z-decomp} and \eqref{find_R'} can be expressed in the matrix form
\begin{equation}\label{find_R'_X'}
    \mathcal{A}\cdot
    \begin{pmatrix}
        R'\\
        X'
    \end{pmatrix} = 
    \begin{pmatrix}
        -1 + \Omega - K - x + y - \frac{p^2 f_X}{6\kappa^2 f_R} - \frac{f_{R\phi}}{f_R}\frac{p}{\kappa} \\
        -6Xf_X^2 - 12Xf_X\frac{\kappa}{p}\frac{f_{\phi}}{R}y - 2Xf_X f_{X\phi}\frac{p}{\kappa}\,,
    \end{pmatrix}
\end{equation}
where the matrix $\mathcal{A}$ is
\begin{equation}\label{A}
    \mathcal{A} =
    \begin{pmatrix}
        \frac{f_{RR}}{f_R} & \frac{f_{RX}}{f_R}\\
        2Xf_X f_{XR} & f_X^2 + 2Xf_X f_{XX}
    \end{pmatrix}\,.
\end{equation}
As long as the matrix $\mathcal{A}$ is non-degenerate, it is possible to find $R'=R'(x,y,q,p,\Omega,K),\,X'=X'(x,y,q,p,\Omega,K)$, and hence $g_1=g_1(x,y,q,p,\Omega,K),\,g_2=g_2(x,y,q,p,\Omega,K)$ by solving the system \eqref{find_R'_X'}. This allows obtaining an autonomous dynamical system.

Provided the dynamical system \eqref{DS_reduced} is regular, the success of this formulation still depends on two crucial steps:
\begin{itemize}
    \item First, it should be possible to explicitly solve for $R=R(x,y,q,p)$ from Eq.\eqref{R}.
    \item Provided that the above condition is possible, the matrix \eqref{A} should be non-degenerate so that it is possible to solve for $R'=R'(x,y,q,p,\Omega,K),\,X'=X'(x,y,q,p,\Omega,K)$.
\end{itemize}
The problem is that either or both of these conditions are violated in many of the physically interesting models of the $f(R,\phi,X)$ class, which renders the above procedure rather useless to investigate the phase space of the respective models.

\section{Dynamical system of $f(R,\phi,X)$ gravity: 2nd formulation}\label{Sec: DS_2}

In view of the limitations of the particular dynamical system formulation presented in Sec.\ref{Sec: DS_1}, in this section we present a slightly different formulation, given by a different choice of the $x$ variable. The  new set of dimensionless variables is
\begin{eqnarray}
\label{variables1}
\Omega  = \frac{\kappa^2\rho}{3f_RH^2}, \quad  
x = \frac{1}{6\kappa^2 H^2} ,\quad
y = \frac{R}{6H^2}, \quad 
z = \frac{ {f_R}'}{f_R}, \quad \nonumber \\
K = \frac{k}{a^2H^2}, \quad
q = \kappa\phi, \quad
p = \kappa\phi' = \frac{\kappa\dot\phi}{H},
\end{eqnarray}
The variables $(K,x,y,z)$ are clearly and unambiguously related to the natural dynamical variables $(a,H,R,\dot R)$ for any function $f(R,\phi,X)$, in contrast with the previous set of variables. Moreover, the variable $x$ is, by construction, non-vanishing for \emph{regular} cosmic evolutions, as long as singular situations with $H\to\infty$ are kept out of consideration. Instead of (\ref{R}), for the new variables, we have the relation
\begin{equation}
\label{ellR}
\kappa^2 R = \frac{y}{x},
\end{equation}
which is well defined in the entire phase for regular cosmological evolutions ($x\neq0$) in any  $f(R,\phi,X)$ theory. The new variable $x$ satisfies the equation
\begin{equation}
x' = -\frac{1}{3\kappa^2 H^2} \frac{H'}{H}.
\end{equation}
In terms of the variable (\ref{variables1}), our dynamical system is composed of the following dynamical equations
\begin{subequations}\label{DS_new}
  \begin{eqnarray}
x' &=& -2x(y-K-2)\,, \\
y' &=& y\left[g_2 - 2(y-K-2) \right]\,, \\
z' &=& 1 - 3{\frac{\kappa^2 f}{f_R} }x + y + K - 3w\Omega - z(y + z - K)\,,\\
K' &=& -2K (y-K-1)\,, \\
\Omega' &=& \Omega(1 - 3w - 2y - z + 2K)\,, \\
q' &=& {p}\,, \\ 
p' &=& -\left(\frac{f_{XR}}{f_X}R' + \frac{f_{XX}}{f_X}X'\right)p - \frac{f_{X\phi}}{f_X}p^2 -\left(y+1-K\right) p - \frac{6{\kappa^3} f_\phi}{f_X}x\,. \\
\end{eqnarray}  
\end{subequations}
along with the Friedmann constraint
\begin{equation}
    z = -1 + \Omega - K - \frac{\kappa^2 f}{f_R}x + y - \frac{f_X p^2}{6\kappa^2 f_R}.\label{constnew}
\end{equation}
In the above
\begin{equation}
  g_2=\frac{R'}{R}.
\label{g2}
\end{equation}
The $p'$-equation above can also be conveniently expressed as
\begin{equation}\label{fxp'_new}
    (p f_X)' = -\left(y +1 -K \right)p f_X - 6{\kappa^3}x f_\phi . 
\end{equation}

Like before, in arriving at the system \eqref{DS_new}, we have assumed that $f_R\neq0$ and $f_X\neq0$ identically. The justifiability of these assumptions follows the same line of discussions as for the earlier formulation. The dynamical system \eqref{DS_new} can still become singular \emph{locally} if $R$ or $f_R$ or $f_X$ vanish \emph{locally}; such physical situations will not be captured by this particular dynamical system formulation. One slight advantage of this formulation over the previous formulation is that this system remains regular even in physical situations with $f=0$ locally, which was not the case for the first formulation.

As before, choosing to eliminate $z$ using the constraint \eqref{constnew}, we arrive at the reduced dynamical system
\begin{subequations}\label{DS_new_reduced}
   \begin{eqnarray}
x' &=& -2x(y-K-2)\label{x'new} \,,\\
y' &=& y\left(g_2 - 2(y-K-2) \right)\label{y'new}\,, \\
\Omega' &=& \Omega\left[-4 - 3w - \Omega + \frac{\kappa^2 f}{f_R}x + \frac{f_X p^2}{6\kappa^2 f_R} - 3(y-K-2)\right]\,, \label{Omega'new}\\
K' &=& -2K (y-K-1)\,, \label{K'new} \\
q' &=& {p} \,,\label{q'new}  \\ 
p' &=& -\left(\frac{f_{XR}}{f_X}R' + \frac{f_{XX}}{f_X}X'\right)p - \frac{f_{X\phi}}{f_X}p^2 -\left(y+1-K\right) p - \frac{6{\kappa^3} f_\phi}{f_X}x\,.\label{p'new}
\end{eqnarray} 
\end{subequations}

Taking into account (\ref{ellR}) and that
\begin{equation}
\label{Xell}
X = -\frac{p^2}{12\kappa^4 x},
\end{equation}
one can write $f(R,\phi,X)$ and all its partial derivatives appearing in the equations (\ref{constnew})-(\ref{p'new}) as
functions of the dynamical variables $(x,y,q,p)$ for \emph{any} given functional form $f(R,\phi,X)$. 

The only thing we are now left with before we can make the system \eqref{DS_new_reduced} autonomous is $g_2=\frac{R'}{R}$. 
From the definition of $z$, we have
\begin{equation}
\label{zz}
z = \frac{{f_R}'}{f_R} = \frac{f_{RR}}{f_R}R' + \frac{f_{R\phi}}{f_R}\frac{p}{\kappa} + \frac{f_{RX}}{f_R}X' = -1 + \Omega - K - \frac{\kappa^2 f}{f_R}x + y - \frac{f_X p^2}{6\kappa^2 f_R}\,.
\end{equation}
On the other hand, taking derivative of $X=-\frac{1}{2}\dot{\phi}^2$ and using Eq.\eqref{p'new}, we arrive at
\begin{equation}\label{find_X'}
\left(\frac{p\,f_{XX}}{f_X}-\frac{6\kappa^4 x}{p}\right)X' = -3p - \frac{f_{X\phi}}{f_X}p^2 - \frac{6\kappa^3 f_{\phi}}{f_X}x - \left(\frac{p\,f_{XR}}{f_X}\right)R'\,.
\end{equation}
The two equations \eqref{zz} and \eqref{find_X'} can be expressed in the matrix form
\begin{equation}\label{find_R'_X'_new}
    \mathcal{B}\cdot
    \begin{pmatrix}
        R'\\
        X'
    \end{pmatrix} = 
    \begin{pmatrix}
        -1 + \Omega - K - \frac{\kappa^2 f}{f_R}x + y - \frac{f_X p^2}{6\kappa^2 f_R} - \frac{f_{R\phi}}{f_R}\frac{p}{\kappa} \\
        -3p - \frac{f_{X\phi}}{f_X}p^2 - \frac{6\kappa^3 f_{\phi}}{f_X}x
    \end{pmatrix}\,,
\end{equation}
where the matrix $\mathcal{B}$ is
\begin{equation}\label{B}
    \mathcal{B} =
    \begin{pmatrix}
        \frac{f_{RR}}{f_R} & \frac{f_{RX}}{f_R}\\
        \frac{p\,f_{XR}}{f_X} & \frac{p\,f_{XX}}{f_X}-\frac{6\kappa^4 x}{p}
    \end{pmatrix}\,.
\end{equation}
As long as the matrix $\mathcal{B}$ is non-degenerate, it is possible to find $R'=R'(x,y,q,p,\Omega,K)$, and hence $g_2=g_2(x,y,q,p,\Omega,K)$ by solving the system \eqref{find_R'_X'_new}. This allows obtaining an autonomous dynamical system.

Two distinct advantages of the current dynamical system formulation over the previous formulation is that
\begin{itemize}
    \item In the current formulation, the dynamical system is regular even when $f=0$.
    \item It is always possible to express $R$ and $X$ using dynamical variables (Eqs.\eqref{ellR} and \eqref{Xell}); there is not invertibility requirement as in Eq.\eqref{R}.
\end{itemize}
The latter requirement, in particular, significantly extends the applicability of the formulation, as such invertibility relations are a big hurdle in forming an autonomous system. We must admit that the condition for non-degeneracy of the matrix $\mathcal{B}$ is still there.

\section{Application of the first formulation: A nonminimal derivative coupling model}\label{Sec: NMDC}

As an application of the first dynamical system formulation presented in Sec.\ref{Sec: DS_1}, let us apply the formulation to investigate the solution space of a nonminimal derivative coupling (NMDC) model, given by the Jordan frame action
\begin{equation}\label{NMDC_action}
    \mathcal{S} = \int d^4 x \sqrt{-g} \left[ \frac{1}{2\kappa^2}F(\phi)R - \frac{1}{2}\partial_{\alpha}\phi \partial^{\alpha}\phi - \frac{1}{2}R\,h(\phi)\partial_{\alpha}\phi \partial^{\alpha}\phi + \mathcal{L}_m\right]\,,
\end{equation}
where the function $F(\phi)$ is dimensionless and the function $h(\phi)$ is of the dimension $[M]^{-2}$. The action \eqref{NMDC_action} contains, as a special case, the NMDC actions considered in the \cite{Capozziello:1999uwa}.

For the convenience of the reader, let us recall the definitions of the dynamical variables used in the first dynamical system formulation
\begin{eqnarray}
\Omega  = \frac{\kappa^2\rho}{3f_RH^2}, \quad  
x = \frac{f}{6f_R H^2} ,\quad
y = \frac{R}{6H^2}, \quad 
z = \frac{{f_R}'}{f_R} = \frac{\dot{f_R}}{H f_R}, \quad \nonumber \\
K = \frac{k}{a^2H^2}, \quad
q = \kappa\phi, \quad
p = \kappa\phi' = \kappa\frac{\dot{\phi}}{H} .
\end{eqnarray}

Comparing with the action \eqref{action}, we obtain
\begin{equation}
    f(R,\phi,X) = \left[F(\phi) - 2\kappa^2 h(\phi)X\right]R - 2\kappa^2 X = F(\phi)R - 2\kappa^2 \left[1 + h(\phi)R\right]X\,.
\end{equation}
Substituting this into Eq.\eqref{R} and performing some straightforward but tedious algebra, one can solve for $R$ as
\begin{equation}
   R(x,y,q) = \frac{y}{h(q/\kappa)}\left[ \frac{1}{x-y} - \frac{F(q/\kappa)}{p^2/6} \right]\,. 
\end{equation}
where we have used the relation \eqref{X} to replace $X$ with $-\frac{R p^2}{12\kappa^2 y}$, and replaced $\phi=q/\kappa$. Using this one can also obtain
\begin{equation}
    X(x,y,q) = \frac{1}{2\kappa^2 h(q/\kappa)}\left[ F(q/\kappa) - \frac{p^2/6}{x-y} \right]\,.
\end{equation}
Since $R$, and hence $X$, can be uniquely solved in terms of the dynamical variables for the theory \eqref{NMDC_action}, the phase space of this theory can be successfully investigated using the first dynamical system formulation of Sec.\ref{Sec: DS_1}.

The phase space analysis can be done generically for any given $F(\phi)$ and $h(\phi)$. However, in practice, the analytic treatment becomes quite heavy, and for the purpose of showing the applicability of the first dynamical system formulation, we present below the analysis for the simple toy model with $F(\phi)=1$ and $h(\phi)=\kappa^2 \xi$, with $\xi$ being a dimensionless coupling constant. The simplified action becomes
\begin{equation}\label{NMDC_action_simpl}
    \mathcal{S} = \int d^4 x \sqrt{-g} \left[ \frac{1}{2\kappa^2}R - \frac{1}{2}\partial_{\alpha}\phi \partial^{\alpha}\phi - \frac{1}{2}\kappa^2 \xi\,R\,\partial_{\alpha}\phi \partial^{\alpha}\phi + \mathcal{L}_m\right]\,,
\end{equation}
This simplistic toy model still incorporates the NMDC term $-R\,X$ in the Lagrangian; so one can still expect to see the characteristic features of NMDC in the resulting theory.

For the toy model \eqref{NMDC_action_simpl}, one has
\begin{subequations}\label{f_fders_NMDC}
    \begin{eqnarray}
        f &=& (1 - 2\kappa^4 \xi X)R - 2\kappa^2 X = R - 2\kappa^2(1 + \kappa^2 \xi R)X\,,
        \\
        f_R &=& 1 - 2\kappa^4 \xi X\,,
        \\
        f_\phi &=& 0\,,
        \\
        f_X &=& -2\kappa^2 (1 + \kappa^2 \xi R)\,,
        \\
        R &=& \frac{1}{\kappa^2 \xi}\left(\frac{y}{x-y} - \frac{6y}{p^2}\right)\,,
        \\
        X &=& \frac{1}{2\kappa^4 \xi}\left[1 - \frac{p^2}{6(x-y)}\right]\,.
    \end{eqnarray}
\end{subequations}
Substituting everything into the system of equations given by Eq.\eqref{find_R'_X'} and performing some straightforward calculations, one can solve for $X'$ and $R'$:
\begin{subequations}
    \begin{eqnarray}
        X' &=& -\frac{p^2 z}{12\kappa^4 \xi(x-y)}\,,
        \\
        R' &=& -\frac{1}{\kappa^2 \xi}\left[3 + 3y\frac{p^2 - 6(x-y)}{p^2(x-y)} + \frac{1}{2}\frac{p^2 z}{p^2 -6(x-y)} + \frac{1}{2}\frac{p^2 y z}{p^2(x-y)}\right]\,,
    \end{eqnarray}
\end{subequations}
This allows us to obtain the following autonomous dynamical system (assuming pressureless matter and spatial flatness motivated by the late-time cosmology)
\begin{subequations}\label{NMDC_DS_1}
\begin{align}
        x' &=  \frac{\left(p^2 x + 6 y (y - x)\right)
        \left(p^4 (x + y + \Omega - 7)
            - 12 p^2 (x - y)(x + 2y + \Omega - 4)
            + 144 y (x - y)^2\right)}{2p^2 \left(p^2 - 6x + 6y\right)^2}\nonumber
            \\
&\quad + \frac{12 x y (x - y)}{p^2} - x(x+3y+\Omega-5)\,,
\\
y' &= -\frac{\left(p^2 x+6 y (y-x)\right) \left(p^2 (x+y+\Omega +5)-12 (y+3) (x-y)\right)}{2 \left(p^2-6 x+6 y\right)^2}-2 (y-2) y\,,
\\
\Omega' &= \frac{12 y \Omega  (x-y)}{p^2}-\Omega  (x+3 y+\Omega -2)\,,
\\
q' &= p\,,
\\
p' &= \frac{p^3 (x-y+\Omega +3)+24 p (y-x)}{2 \left(p^2-6 x+6 y\right)}\,.
\end{align}
\end{subequations}

The system \eqref{NMDC_DS_1}, although autonomous, is singular at $p=0$ and $p^2=6(x-y)$, with poles of order two. The dynamical system can be regularized by redefining the time variable as
\begin{equation}\label{time_redef}
    N\to\tilde{N}:d\tilde{N}=\frac{dN}{p^2 (p^2 -6x+6y)^2}\,.
\end{equation}
The time redefinition is chosen keeping in mind the order of the poles such that the singularity of the system is exactly regularized while no artificial spurious fixed points, either regular or singular, is introduced (see, e.g., \cite[Sec.3]{Bouhmadi-Lopez:2016dzw}). With respect to $\tilde{N}$, the regularized dynamical system becomes
\begin{subequations}\label{NMDC_DS_2}
\begin{align}
        x' &=  
    \frac{\left(p^2 x + 6 y (y - x)\right)
        \left(
            p^4 (x + y + \Omega - 7)
            - 12 p^2 (x - y)(x + 2y + \Omega - 4)
            + 144 y (x - y)^2
        \right)}{2} \\\nonumber
&\quad +12xy(x-y)(p^2-6x+6y)^2 - x(x+3y+\Omega-5)p^2 (p^2-6x+6y)^2\,,
\\
y' &= -p^2\left[\frac{\left(p^2 x+6 y (y-x)\right) \left(p^2 (x+y+\Omega +5)-12 (y+3) (x-y)\right)}{2} + 2y(y-2)(p^2-6x+6y)^2\right]\,,
\\
\Omega' &= \Omega\left[p^2 - 6(x-y)\right]\left[12y(x-y) - p^2(x+3y+\Omega-2)\right]
\\
q' &= p^3[p^2-6(x-y)]^2\,,
\\
p' &= \frac{1}{2} p^3 \left[p^2 - 6(x-y)\right]\left[p^2 (x-y+\Omega +3) - 24(x-y)\right]\,.
\end{align}
\end{subequations}
There are two noteworthy features that can be observed from the system \eqref{NMDC_DS_2}.
\begin{itemize}
    \item The $q$-equation completely decouples. $q$ does not appear in any other dynamical equation. The scalar field affects the cosmological dynamics only through its velocity and not its absolute value. This is possibly a consequence of the absence of any potential.
    \item Another interesting feature is the disappearance of the dimensionless coupling strength parameter $\xi$, indicating that the qualitative features of the cosmological dynamics are independent of its strength.
\end{itemize}

\subsection{Invariant submanifold analysis}

It is instructive to study the invariant submanifolds of the system \eqref{NMDC_DS_2}. All solutions starting on these submanifolds will always be constrained to remain on them. Mathematically, system \eqref{NMDC_DS_2} presents seven clear invariant submanifolds:
\begin{itemize}
\item $p=0$: This corresponds to the actual physical situation when the field freezes.
\item $p=\pm\sqrt{6(x-y)}$: Strictly speaking, these two invariant submanifolds physically exist only if $x>y$. Taking into account the definitions of the dynamical variables in Eq.\eqref{variables}, the condition $x>y$ implies
\begin{equation}
    \frac{f-Rf_R}{f_R}>0\,.
\end{equation}
For the particular action \eqref{NMDC_action_simpl}, the above condition gives
\begin{equation}
    \frac{-2\kappa^2 X}{1-2\kappa^4 \xi X}=\frac{\kappa^2 \dot{\phi}^2}{1 + \kappa^4 \xi \dot{\phi}^2}>0\,.
\end{equation}
The above condition can be \emph{guaranteed} for all values of $\dot{\phi}$ if $\xi>0$. For $\xi<0$, the requirement for the existence of the invariant submanifolds $p=\pm\sqrt{6(x-y)}$ restricts the speed of the scalar field as $|\dot{\phi}|<\frac{1}{\kappa^2 \sqrt{|\xi|}}$.

From the $p'$-equation in \eqref{NMDC_DS_2} it might appear that the field also freezes on the invariant submanifolds $p=\pm\sqrt{6(x-y)}$, which leads to an apparent confusion as $p=\frac{dq}{dN}\neq0$. One needs to be careful about the interpretation here, as we have used a \emph{velocity-dependent} time redefinition \eqref{time_redef} on the phase space. The relation $p=\frac{dq}{dN}$ is defined with respect to the logarithmic time $\log(a)$, whereas the $q'$-equation in \eqref{NMDC_DS_2} is the derivative of $q$ with respect to the redefined time variable $\tilde{N}$. The apparent \emph{mathematical freezing} of the field at $p=\pm\sqrt{6(x-y)}$ is due to this velocity-dependent time redefinition; physically, there is no freezing of the scalar field here. This acts as a reminder to be careful about physical interpretations of the phase space structure when a time-redefinition is involved.
\item $p=\sqrt{\frac{24(x-y)}{x-y+\Omega+3}}$: These two invariant submanifolds physically exist only if
    \begin{equation}\label{existence}
       \frac{x-y}{x-y+\Omega+3}>0\,.
    \end{equation}
Note that the above condition can be \emph{guaranteed} for $x>y$. We have seen earlier that the condition $x>y$ can \emph{always} be met for $\xi>0$. Notice that $\xi>0$ also makes $f_R>0$ identically, so that $\Omega>0$. Therefore, we can say that taking $\xi>0$ guarantees the existence of the invariant submanifold $p=\pm\sqrt{6(x-y)}$. 
\item $x-y=0$: The existence of this hidden invariant submanifold can be made apparent if one makes a variable transformation $x\to\tilde{x}=x-y$. Then, from the $x'$-equation and the $y'$-equation of the system \eqref{NMDC_DS_2} one can write
\begin{equation}
    \tilde{x}' = -2\,\tilde{x}\left(p^2 - 6\tilde{x}\right)\left[p^4 (y-2) - 3\,p^2 \tilde{x}(\tilde{x}+4y+\Omega-5) + 36\,y\,\tilde{x}^2\right]\,.
\end{equation}
Physically, this corresponds to the case $R\,f_R=f$, which, for the Lagrangian \eqref{NMDC_action_simpl}, implies $X=0$. Therefore, for the particular NMDC model \eqref{NMDC_action_simpl} where there is no scalar field potential, the invariant submanifold $x-y=0$ bears no distinct physical meaning from the invariant submanifold $p=0$. The latter conclusion would not be true in the presence of a scalar field potential.
\item $\Omega=0$: There is, of course, the invariant submanifold $\Omega=0$, containing all the vacuum solutions.
\end{itemize}

Notice that all seven invariant submanifolds presented above are characterized by one constraint. Therefore, all of them are 4-dimensional invariant submanifolds, each of which divides the entire phase space into two disjoint sectors. This implies the non-existence of any global past or future attractors.

It is imperative to characterize the invariant submanifolds in terms of their dynamical stability. It is easy to understand the stability of the invariant submanifold $p=0$ by expanding the right-hand side of the $p'$-equation of \eqref{NMDC_DS_2} near $p=0$:
\begin{equation}
    p' \approx 72(x-y)^2 p^3\,,
\end{equation}
which shows that the invariant submanifold $p=0$ is always unstable. The stability of the other invariant submanifolds depends on the values of $x,y,\Omega$. 

The unstable nature of the $p=0$ invariant submanifold is an interesting feature, which means that the field does not tend to freeze. Typically, when there is a potential with a minimum, the field tends to settle down there, leading to its freezing. The action under consideration \eqref{NMDC_action_simpl} does not have a potential, and hence there is no scope for the field to freeze.

Although we have been able to figure out some interesting results with the invariant submanifold analysis in this section, let us put a disclaimer that they are based on the simplified toy NMDC action \eqref{NMDC_action_simpl}. The toy action \eqref{NMDC_action_simpl} is limited and exploratory in nature, purely serving to show the applicability, or rather limitations thereof, of the first dynamical system formulation. For a generic NMDC theory, not only can the functions $F(\phi)$ and $h(\phi)$ be variables, but there can be other possible coupling terms like $h(\phi)R^{\mu\nu}\partial_{\mu}\phi$ or $g(\phi)R$ in the action.

\subsection{Fixed point analysis}

There are four distinct families of fixed points to be found corresponding to the system \eqref{NMDC_DS_2}:
\begin{itemize}
    \item $\mathcal{F}_1\equiv(x,0,\Omega,q,0)$
    \item $\mathcal{F}_2\equiv\{(x,y,\Omega,q,0): x=y\}$.
    \item $\mathcal{F}_{3\pm}\equiv\{(x,y,\Omega,q,p): \Omega=1-x+y,\,p=\pm\sqrt{6(x-y)}\}$
\end{itemize}
Each of the families of fixed points is characterized by two constraints in a 5-dimensional phase space, and hence is a 3-parameter family of fixed points. They are listed in Table \ref{tab:NMDC_finite_fp}. The three families of fixed points intersect at the line $\mathcal{L}_q\equiv(0,0,1,q,0)$.
\begin{table}[H]
\centering
\resizebox{\textwidth}{!}{
\begin{tabular}{|c|c|c|c|}
\hline
F.P. & $(x,y.\Omega,q,p)$ & Stability & Cosmology
\\ \hline
$\mathcal{F}_1$ & $(x^*,0,\Omega^*,q^*,0)$ & Non-hyperbolic & $a(t)\sim t^{1/2}$
\\ \hline
$\mathcal{F}_2$ & $(y^*,y^*,\Omega^*,q^*,0)$ & Non-hyperbolic & Depends on the value of $y^*$
\\ \hline
$\mathcal{F}_{3\pm}$ & $\left(x^*,y^*,1-x^*+y^*,q^*,\pm\sqrt{6(x^*-y^*)}\right)$ & Non-hyperbolic & Depends on the value of $y^*$
\\ \hline
\end{tabular}
}
\caption{Finite fixed points of the dynamical system \eqref{NMDC_DS_2}. There are three distinct families of fixed points, all of which are 3-parameter families.}
\label{tab:NMDC_finite_fp}
\end{table}

Unfortunately, all the fixed points come out to be non-hyperbolic, meaning their stability investigation is beyond the normal linear stability analysis. Without employing more sophisticated treatments like center manifold analysis, no conclusive statements about attractors, instabilities, or any global behaviour can be made. We do not go further into the stability analysis of the fixed points in this paper, and argue this as a possible limitation of the first dynamical system formulation that we have applied to study this particular model.

\section{Application: Higgs-$R^2$ inflationary model}\label{Sec: SH}

As an application of the second dynamical system formulation presented in Sec.\ref{Sec: DS_2}, let us apply the formulation to investigate the solution space of the $R^2$-Higgs model of inflation \cite{He:2018gyf}\footnote{The possibility of an $R^2$-Higgs inflation was also explored earlier in \cite{Salvio:2015kka}.}. The model is given by the Jordan frame action
\begin{equation}\label{SH_action}
    \mathcal{S} = \int d^4 x \sqrt{-g} \left[ \frac{1}{2\kappa^2}(R + \alpha R^2) + \frac{1}{2}\xi\phi^2 R - \frac{1}{2}\partial_{\alpha}\phi \partial^{\alpha}\phi - \frac{\lambda}{4}\phi^4 \right]\,,
\end{equation}
This is essentially a two scalar field inflationary model, with the nonminimally coupled Higgs field $\phi$ arising explicitly in the Lagrangian, whereas the $R^2$ term giving rise to an effective scalaron field that can be made apparent by transforming to the Einstein frame description. The analysis in the original paper \cite{He:2018gyf} was performed in the Einstein frame. In this work, however, we stick to the Jordan frame.

Comparing with the action \eqref{action}, we obtain
\begin{equation}
    f(R,\phi,X) = \left[(1+\kappa^2\xi\phi^2)R + \alpha R^2\right] - 2\kappa^2 \left(X + \frac{\lambda}{4}\phi^4\right)\,.
\end{equation}
The inflationary model as proposed in Ref.\cite{He:2018gyf} was formulated on the spatially flat FLRW cosmology ($K=0$). Moreover, since this is an inflationary model, there is no matter fluid present in the background ($\Omega=0$). The phase space of the model is therefore 4-dimensional. We choose the variables $x,y,q,p$ to span the phase space, whereas the variable $z$ is determined by the Friedmann constraint.

Firstly, let us show that the first dynamical system formulation presented in Sec.\ref{Sec: DS_1} fails to analyze the phase space of the model given by the Lagrangian \eqref{SH_action}. Recall that the success of the first dynamical system formulation hinges on the invertibility of the relation \eqref{R} to obtain $R=R(x,y,q,p)$. For the particular model under consideration, the relation \eqref{R} gives
\begin{equation}
    \frac{y}{x} \equiv \frac{R f_R}{f} = \frac{R(1 + \kappa^2\xi\phi^2 + 2\alpha R)}{[(1+\kappa^2\xi\phi^2)R+\alpha R^2]-2\kappa^2 X-\frac{\kappa^2\lambda}{2}\phi^4}\,,
\end{equation}
which can be written as
\begin{equation}
\left(2-\frac{y}{x}\right)\alpha R^2 + \left[(1+\kappa^2\xi q^2)\left(1-\frac{y}{x}\right)-\frac{p^2}{6x}\right]R + \frac{\lambda}{2\kappa^2}\frac{y}{x}q^4=0\,.
\end{equation}
The above relation is not invertible. One cannot obtain a unique expression for $R=R(x,y,q,p)$ from the above equation. Therefore, the first dynamical system formulation, as formulated in Sec.\ref{Sec: DS_1}, fails to analyze the solution space of this model.

On the contrary, there is no invertibility requirement limiting the applicability of the second dynamical system formulation presented in Sec.\ref{Sec: DS_2}, which, as we show below, can successfully analyze the model \eqref{SH_action}. 

For the convenience of the reader, let us recall the definitions of the dynamical variables used in the second dynamical system formulation
\begin{eqnarray}
\Omega  = \frac{\kappa^2\rho}{3f_RH^2}, \quad  
x = \frac{1}{6\kappa^2 H^2} ,\quad
y = \frac{R}{6H^2}, \quad 
z = \frac{ {f_R}'}{f_R}, \quad \nonumber \\
K = \frac{k}{a^2H^2}, \quad
q = \kappa\phi, \quad
p = \kappa\phi' = \frac{\kappa\dot\phi}{H},
\end{eqnarray}

For the model under consideration, one can calculate the following quantities in terms of the dynamical variables
\begin{subequations}\label{f_fders_SH}
    \begin{eqnarray}
        f &=& \frac{1}{\kappa^2}\frac{y}{x}\left(1 + \xi q^2 + \frac{\alpha}{\kappa^2}\frac{y}{x}\right) + \frac{p^2}{6\kappa^2 x} - \frac{\lambda}{2\kappa^2}q^4\,,
        \\
        f_R &=& 1 + \xi q^2 + 2\frac{\alpha}{\kappa^2}\frac{y}{x}\,,
        \\
        f_\phi &=& \frac{2}{\kappa}q\left(\xi \frac{y}{x} - \lambda q^2\right)\,,
        \\
        f_X &=& -2\kappa^2\,.
    \end{eqnarray}
\end{subequations}
One can also use Eqs.\eqref{zz} and \eqref{find_X'} to obtain
\begin{subequations}
    \begin{eqnarray}
       X' &=& \frac{1}{2\kappa^4}p \left[\frac{p}{x} - 2q \left(\xi\frac{y}{x} - \lambda q^2\right)\right]\,,
       \label{Xprime}
       \\
       R' &=& \frac{z}{2\alpha}\left(1 + \xi q^2 + 2\frac{\alpha}{\kappa^2}\frac{y}{x}\right) - \frac{\xi}{\alpha} p\,q\,,
    \label{Rprime}
    \end{eqnarray}
\end{subequations}
A crucial point to note here is that, while one can write Eqs.\eqref{f_fders_SH} and Eq.\eqref{Xprime} in the limit $\alpha\to0$, Eqs.\eqref{Rprime}, at least in the way it is written, is valid only if $\alpha\neq0$.

Eq.\eqref{Rprime} gives the expression for $g_2$
\begin{equation}
    g_2\equiv\frac{R'}{R} = z + \frac{\kappa^2}{\alpha}\frac{x}{y}\left[\frac{1}{2}(1+\xi q^2)z - \xi\,p\,q\right]
\end{equation}
Finally, we have the Friedmann constraint \eqref{constnew}
\begin{equation}\label{constr_HS}
    \left[(1 + \xi q^2)x + 2\frac{\alpha}{\kappa^2}y\right](1+z) = \frac{\alpha}{\kappa^2}y^2 + \frac{1}{6}p^2 x + \frac{1}{2}\lambda x^2 q^4\,,
\end{equation}
which can also be expressed as
\begin{equation}
    g_2 = -1 + \frac{1}{2}y - \frac{\kappa^2}{12\alpha}\frac{x}{y}(6 - p^2 + 6\xi q^2 + 12\xi p q - 3\lambda q^4 x)\,.
\end{equation}
We can then use this to eliminate $z$ and write a closed system in terms of $x, y, q, p$. The final dynamical system is as follows:
\begin{subequations}\label{SH_DS_2}
    \begin{eqnarray}
        && x' = -2 x (y-2)\,,
        \\
        && y' = 3y - \frac{3}{2}y^2 - \frac{\kappa^2}{12\alpha}x(6 - p^2 + 6\xi q^2 + 12\xi p q - 3\lambda q^4 x)\,,
        \\
        && q' = p\,,
        \\
        && p' = - p (1+y) - 6\lambda  q^3 x + 6\xi q y\,.
    \end{eqnarray}
\end{subequations}
One can notice that the system \eqref{SH_DS_2} is regular everywhere in the phase space, and is symmetric under the transformation $\{p\to-p,\,q\to-q\}$. The symmetry is an artifact of the symmetry of the scalar field potential. 

It must also be pointed out that in arriving at the regular dynamical system \eqref{SH_DS_2}, we have made the inherent assumption that $x\neq0$ (since $x$ appears in the denominator of Eqs.\eqref{Rprime} and \eqref{Xprime}). In other words, although the system \eqref{SH_DS_2} appears to be regular in the limit $x\to0$, the physical domain of validity of the system is given by $x>0$, and hence $x=0$ lies at the asymptotic boundary of this domain. So, strictly speaking, the system \eqref{SH_DS_2} describes the phase space as long as the cosmology is regular.

We will present below two limit cases of \eqref{SH_action} for which the number of independent degrees of freedom changes. For both these cases, the dynamical system reduces to two dimensions; see \cite{Jarv:2021qpp} and \cite{Barrow:2006xb} respectively. They should therefore be understood as constrained reductions of the full system, not as ordinary substitutions in the four-dimensional autonomous system \eqref{SH_DS_2}. In particular, the pure Higgs limit $\alpha=0$ removes the scalar degree of freedom, while the pure $R^2$ limit corresponds to restricting the Higgs field to the invariant sector $q=p=0$. The purpose of these reductions is to check that the general formalism reproduces the known lower-dimensional phase spaces once the appropriate constraints are imposed.

\subsection{Constrained Higgs-inflation limit $(\alpha=0)$: Comparison with Jarv et.al. \cite{Jarv:2021qpp}}

The pure Higgs inflation model is obtained by setting $\alpha=0$, which removes the scalaron degree of freedom. Since the full four-dimensional system \eqref{SH_DS_2} was obtained using relations that contain $1/\alpha$, this limit cannot be taken by direct substitution in Eq.~(55). Instead, one must return to the constraint equations and impose $\alpha=0$ before eliminating the dependent variables. The resulting reduced phase space is two-dimensional and may be parametrized by $(q,p)$. The variables $x,y,z$ are then not independent dynamical variables, but functions of $(q,p)$ determined by these constraints. Substituting $f_R=1+\xi\kappa^2\phi^2=1+\xi q^2$ into the definition $z=\frac{f_R'}{f_R}$ and using $q'=p$, one gets an additional constraint
\begin{equation}\label{z(qp)}
    z = \frac{2\xi q p}{1 + \xi q^2}\,.
\end{equation}
Substituting the expression for $z$ from \eqref{z(qp)} into the Friedmann constraint \eqref{constr_HS}, and keeping in mind that $\alpha=0,\,x>0$, one can solve for $x$
\begin{equation}\label{x(qp)}
    x = \frac{6(1 + \xi q^2) + 12\xi q p - p^2}{3\lambda q^4}\,.
\end{equation} 
We need an additional constraint relating $y$ to $q,\,p$. To achieve this goal, we proceed as follows. First, setting $\alpha=0$ and considering that $x>0$ in the Friedmann equation \eqref{constr_HS}, we take its derivative. This gives us
\begin{equation}\label{z'_SH_1}
    z' = \frac{-p^2 (y+1)+6 \xi  p q (y-z-1)-3 \lambda  q^4 x (y-2)}{3(1 + \xi  q^2)}\,.
\end{equation}
Next, notice that $z'$ can also be calculated straightaway by taking the derivative of the expression \eqref{z(qp)}, and using the $\{q',p'\}$-equations from \eqref{SH_DS_2}.
\begin{equation}\label{z'_SH}
    z' = -(1+24\xi)\left(\frac{2\xi q p}{1 + \xi q^2}\right) - 2\xi q y\left(\frac{p - 6\xi q}{1 + \xi q^2}\right) - 24\xi + \frac{2\xi p^2(3 + \xi q^2)}{(1 + \xi q^2)^2}\,.
\end{equation}
We equate these two expressions of $z'$ and substitute the solved expressions for $x(q,p)$ and $z(q,p)$ from Eqs.\eqref{x(qp)} and \eqref{z(qp)} respectively. The resulting algebraic relation allows us to solve for $y(q,p)$
\begin{equation}\label{y(qp)}
    y = \frac{(6 \xi +1) \left(-p^2+8 \xi  p q+4 \xi  q^2+4\right)}{2 \left(6 \xi ^2 q^2+\xi  q^2+1\right)}
\end{equation}
Substituting the expressions for $x(q,p),\,y(q,p),\,z(q,p)$ from the equations \eqref{x(qp)},\eqref{y(qp)} and \eqref{z(qp)} into the $p'$-equation of the system \eqref{SH_DS_2} gives us the 2-dimensional dynamical system corresponding to the pure Higgs inflation scenario
\begin{subequations}\label{H_DS}
    \begin{eqnarray}
        q' &=& p\,,
        \\
        p' &=& -\frac{p}{q}(5p+3q) + \frac{6 \xi  q \left(p^3-12 p-4 q\right)+p^2 (p q+14)-24}{2q(1 + \xi q^2 + 6\xi^2 q^2)}\,.
    \end{eqnarray}
\end{subequations}
Thus, finally, Eq. \eqref{H_DS} is the flow induced by the full equations on the constrained Higgs-inflation phase space; it is not obtained by setting $\alpha=0$ directly in the four-dimensional \eqref{SH_DS_2}. 
Although different in looks, one can verify that the 2-dimensional dynamical system \eqref{H_DS} is exactly the same as obtained by Toporensky and Jarv \cite[Eq.(54)]{Jarv:2021qpp}, who had earlier considered the phase space of nonminimally coupled inflaton using the same variables $q,\,p$ that we have considered here. The fact that we have been able to retrieve the same dynamical system from the generic dynamical system framework of $f(R,\phi,X)$ gravity serves as a consistency check of our framework.

\subsection{Invariant $R^2$-inflation sector $(p=q=0)$: Comparison with Barrow et. al. \cite{Barrow:2006xb}}

Reducing the case to pure quadratic gravity inflation, known as Starobinsky inflation, is an easier task. The restriction $q=p=0$ is an invariant submanifold of the full system, since
$q'=p=0$ and
$p'=-p(1+y)-6\lambda q^3 x+6\xi q y=0$
whenever $q=p=0$, so that the system \eqref{SH_DS_2} reduces to
\begin{subequations}\label{Staro_DS_2}
\begin{eqnarray}
x' &=& -2x(y-2)\,,
\\
y' &=& 3y-\frac{3}{2}y^2 - \frac{\kappa^2}{2\alpha}x\,.
\end{eqnarray}
\end{subequations}
The phase space of vacuum Bianchi type-I and type-II universe in the generalized quadratic gravity of the form $f(R,R_{\mu\nu}R^{\mu\nu}) = R + \alpha R^2 + \beta R_{\mu\nu}R^{\mu\nu} - 2\Lambda$ has previously been analyzed in details in Ref.\cite{Barrow:2006xb}. The phase space is generically 8-dimensional (see [Eqs.(9)--(18)]\cite{Barrow:2006xb}). If one specializes to the case of a homogeneous isotropic universe with $\beta=\Lambda=0$, one can check that the 8-dimensional phase space of [Eqs.(9)--(18)]\cite{Barrow:2006xb} reduces to a 2-dimensional phase space, which is essentially the same as the Eq.\eqref{Staro_DS_2} above\footnote{For the convenience of the reader, we present that in this case the constraint equation \cite[Eq.(18)]{Barrow:2006xb} simplifies to $Q_2 = \frac{1}{2}Q^2 - 3Q - \frac{1}{4}B$, with their variables $\{B,Q\}$ related to our variables $\{x,y\}$ as $B=2\frac{\kappa^2}{\alpha}x,\,Q=y-2$.}. This is a further consistency check on our generic $f(R,\phi,X)$ dynamical system framework.

\subsection{The large $\xi$ limit}

In \cite{He:2018gyf} it was argued that non-minimal coupling term can be partially regarded as an additional contribution to the Higgs field kinetic term proportional to $\xi$, so that in the limit $|\xi|\gg1$, the original kinetic term of the Higgs field is negligible and the resulting theory is just another quadratic gravity scenario with a modified scalaron mass (see the discussion around \cite[Eq.(4.3)]{He:2018gyf}). No explicit derivation was given in \cite{He:2018gyf}; something which we present in Appendix \ref{app:SH} by analyzing the action \eqref{SH_action} on the FLRW minisuperspace for the reader's convenience.

Here we would like to demonstrate that the argument can also be explained from the dynamical system point of view. Neglecting the original kinetic term of the action in the action is equivalent to taking $p=0$ identically in the dynamical system \eqref{SH_DS_2}, but the field is not necessarily frozen at the potential minimum (i.e. $q\neq0$). Rather, one can solve from the $p'$-equation of the system \eqref{SH_DS_2} that the Higgs field does still have a dynamics, which is implicitly given by the constraint equation
\begin{equation}
    q^2 = \frac{\xi}{\lambda}\frac{y}{x}\,.
\end{equation}
Substituting this value of $q$ into the $y'$-equation one can get
\begin{equation}
    y' = \left(1 - \frac{\kappa^2 \xi^2}{6\alpha\lambda}\right)\left(3y - \frac{3}{2}y^2\right) - \frac{\kappa^2}{2\alpha}x\,.
\end{equation}
Let us now introduce variable changes as
\begin{equation}
    N \to \tilde{N}: d\tilde{N} = dN\left(1 - \frac{\kappa^2\xi^2}{6\alpha\lambda}\right)\,, \qquad x\to\tilde{x}=\frac{x}{\left(1 - \frac{\kappa^2\xi^2}{6\alpha\lambda}\right)}\,.
\end{equation}
This results in a reduced 2-dimensional dynamical system
\begin{subequations}\label{Staro_DS_new_2}
\begin{eqnarray}
\frac{d\tilde{x}}{d\tilde{N}} &=& -2\tilde{x}(y-2)\,,
\\
\frac{dy}{d\tilde{N}} &=& 3y-\frac{3}{2}y^2 - \frac{\kappa^2}{2\alpha}\tilde{x}\,.
\end{eqnarray}
\end{subequations}
The resulting dynamical system \eqref{Staro_DS_new_2}, although written in terms of rescaled $x$ and $N$-variable, is topologically equivalent to the dynamical system of the dynamical system \eqref{Staro_DS_2} for the original quadratic gravity scenario, and therefore have same qualitative characteristics. This is another way of confirming that the limit $|\xi|\gg1$ at which the original kinetic term of the Higgs field can be neglected in the action, reduces effectively to another Starobinsky-like scenario.

\subsection{Invariant submanifold analysis}\label{subsec:inv_sub_SH}

It is instructive to study the invariant submanifolds of the system (\ref{SH_DS_2}). All solutions starting on these submanifolds will always be constrained to remain on them.  The system presents two two-dimensional invariant manifolds\footnote{Technically, we also have a third one-dimensional invariant submanifold, the case corresponding to $x=y=p=0$ and constant $q$, which is a line of fixed points. This will be discussed in the next Section.} :
\begin{itemize}
    \item The first one corresponds to the case $x=y=0$. Since $x=\frac{1}{6\kappa^2 H^2}$ and $y=2+\frac{\dot{H}}{H^2}$, this invariant submanifold contains solutions of the form $a(t)\sim\sqrt{t-t_0}$ ($t>t_0$) at the limit $t\to t_0$ (or $H\to\infty$); namely the big-bang singularity. On this invariant submanifold, one has $q'=p$ and $p'=-p$, {\em i.e.}, the solution $\phi\sim e^{-N}$, irrespective of the value of the parameter $\xi$. This is expected, since the Hubble parameter introduces a friction-like term in the scalar field equation of motion, and the friction term dominates in the large $H$ limit. 
    \item The second invariant manifold is $q=p=0$, corresponding to $\phi=\phi'=0$. This is nothing but the field sitting at the potential minimum. If the field starts at the potential minimum, it remains at the potential minimum.
\end{itemize}

An important comment here is in order. Under Eq.\eqref{SH_DS_2}, we have pointed out that, strictly speaking, the system \eqref{SH_DS_2} describes the cosmological phase space as long as the cosmology is regular ($H\nrightarrow\infty$). In other words, within the domain of physical validity of the system, one has $x>0$. In that sense, the invariant submanifold $x=y=0$, strictly speaking, lies at the asymptotic boundary of the physical phase space represented by the system \eqref{SH_DS_2}.

It is worth investigating the stability of the above invariant submanifolds. To investigate the stability of the invariant submanifold $x=0=y$, we can employ a nice trick based on Gauss' divergence theorem; see \cite{Zaremba} for an earlier analysis and \cite{Furtat} for a more recent exposition. The main idea is that for an autonomous system $\frac{dX}{dN} = F(X(N))$ with $X(N):\mathbb{R}\to \mathbb{R}^n$, one must evaluate the flux of the vector field $F:\mathbb{R}^n\to \mathbb{R}^n$ on a closed simple hypersurface $\delta S$ of $\mathbb{R}^n$, $\Phi = \oint_{\delta S} F\cdot ds$. For the case of positive flux, it is clear that there are solutions escaping the surface $\delta S$, which necessarily implies that its interior $S$ has unstable points. One can say, in this case, that $S\subset \mathbb{R}^n$ is a repulsive subset of the phase space $\mathbb{R}^n$. Notice that one can invoke Gauss' theorem and write $\Phi = \int_S ({\rm div}\, F) \, d{\rm vol}$, and consequently, if we have $ {\rm div}\, F > 0$ in a region $S\subset \mathbb{R}^n$, necessarily $S$ will be a repulsive subset. 

For the system \eqref{SH_DS_2}, one can calculate $ {\rm div}\, F = 6(1-y)$), and hence we cannot have any stable subset in the region $y<1$ of the phase space, irrespective of the values of the parameter $\xi,\,\alpha,\,\lambda$. What this means is that the invariant submanifold $x=0=y$ can never be stable. Notice, however, that this argument says nothing about the region $y>1$. 

The above procedure, unfortunately, does not completely specify the stability of the $q=0=p$ invariant submanifold. What it \emph{does} say, though, is that within the region $y<1$ of the phase space, the invariant submanifold $q=0=p$ cannot be stable; it can be either saddle or unstable. This is an interesting feature, considering that $q=0=p$ corresponds to the minimum of the potential. In the absence of the scalaron, one would expect the Higgs field to roll down the potential and settle at the minimum (e.g. in \cite{Jarv:2021qpp}). However, it appears that if we include a scalaron into the picture, it is \emph{possible} to obtain solution trajectories that push the field away from the minimum. Such trajectories, however, do not contain inflationary solutions, and hence not interesting from the physical point of view. As we will see in the next section, the (super)inflationary solution actually does sit at the minimum of the potential.

\subsection{Fixed point analysis}

The fixed points can be obtained by the following reasoning. Obviously, one should have $p=0$ from (\ref{SH_DS_2}c).
The nontrivial (asymptotic) coordinate $y\equiv R/(6H^2)\equiv \dot H/H^2 + 2 \to 2$ corresponds to a super-inflation fixed point, where $\dot H/H^2 \to 0$ as $t \to \infty$. \footnote{We recall the reader that in a standard power-law scale factor ($ a(t)\sim t^n $) one would have $\dot H/H^2 = -1/n$ and that an exponential inflationary solution (i.e, de Sitter), where $H$ is a constant, yields $\dot H/H^2=0$.}
From that, one is able to obtain  from  (\ref{SH_DS_2}b) and (\ref{SH_DS_2}d) that $x=0$ and $q=0$, which is the point $\mathcal{P}$ in Table \ref{tab:SH_finite_fp}. 
On the other hand, if one assumes $y=0$, then $x=p=0$ but there is no restriction on $q$, which is the "line of fixed points" $\mathcal{L}$ in Table \ref{tab:SH_finite_fp}. 

\begin{table}[H]
\centering
\resizebox{\textwidth}{!}{
\begin{tabular}{|c|c|c|c|c|c|}
\hline
F.P. & $(x,y.q,p)$ & Eigenvalues & Eigenvectors & Stability & Cosmology
\\ \hline
$\mathcal{P}$ & $(0,2,0,0)$ & 
\begin{tabular}{@{}c@{}} 
$-3$ \\ 
$0$ \\ 
$ -\frac{3}{2} \big( 1\pm \sqrt{1+\frac{16}{3} \xi}\big)$\\
\end{tabular} & 
\begin{tabular}{@{}c@{,}c@{,}c@{}}
$\begin{pmatrix} 0 \\ 1 \\ 0 \\ 0 \end{pmatrix}$ &
$\begin{pmatrix} -6\frac{\alpha}{\kappa^2} \\ 1 \\ 0 \\ 0 \end{pmatrix}$ &
$\begin{pmatrix} 0 \\ 0 \\ \frac{3 \pm \sqrt{3} \sqrt{16 \xi +3}}{24 \xi } \end{pmatrix}$ 
\end{tabular} & 
\begin{tabular}{@{}c@{}} 
Saddle for $\xi>-\frac{3}{16}$ \\ 
Nonhyperbolic for $\xi<-\frac{3}{16}$ \\ 
\end{tabular} &
\begin{tabular}{@{}c@{}} 
$a(t)\sim e^{H_0 t^2}$ \\ 
(superinflation) \\ 
\end{tabular}
\\ \hline
$\mathcal{L}$ & $(0,0,q,0)$ & 
\begin{tabular}{@{}c@{}} 
$4$ \\ 
$3$ \\ 
$-1$\\
$0$
\end{tabular} & 
\begin{tabular}{@{}c@{,}c@{,}c@{,}c@{}}
$\begin{pmatrix} -\frac{5 \alpha }{3 q \left(\kappa ^2 \xi +2 \alpha  \lambda  q^2+\kappa ^2 \xi ^2 q^2\right)} \\ \frac{5 \left(\kappa ^2+\kappa ^2 \xi  q^2\right)}{6 q \left(\kappa ^2 \xi +2 \alpha  \lambda  q^2+\kappa ^2 \xi ^2 q^2\right)} \\ \frac{1}{4} \\ 1 \end{pmatrix}$ &
$\begin{pmatrix} 0 \\ \frac{2}{3 \xi  q} \\ \frac{1}{3} \\ 1 \end{pmatrix}$ &
$\begin{pmatrix} 0 \\ 0 \\ -1 \\ 1 \end{pmatrix}$ &
$\begin{pmatrix} 0 \\ 0 \\ 1 \\ 0 \end{pmatrix}$ 
\end{tabular}  & Saddle & 
\begin{tabular}{@{}c@{}} 
$a(t)\sim t$ \\ 
(quasi-FLRW) \\ 
\end{tabular}
\\ \hline
\end{tabular}
}
\caption{Finite fixed points of the dynamical system \eqref{SH_DS_2}. The order in which the eigenvectors are listed follows exactly the order in which the corresponding eigenvalues are listed. It can be noted that $\mathcal{L}$ is a line of saddle fixed points, which explains the appearance of the single zero eigenvalue, whose corresponding eigenvector lies along the $q$-axis. Looking at the eigenvectors, one can notice that in the vicinity of the fixed point $\mathcal{P}$ the total flow is reducible between two orthogonal flows: one component within the ``scalaron plane ($x-y$)'', and another component within the ``Higgs plane ($q-p$)''.}
\label{tab:SH_finite_fp}
\end{table}

Note that the fixed point $\mathcal{P}$ and the line of fixed points $\mathcal{L}$ that we have obtained in Table \ref{tab:SH_finite_fp} are exactly parallel to the fixed points $E$ and $\mathcal{I}$ as obtained by Barrow et.al. \cite{Barrow:2006xb}, and hence bear exactly the same interpretation as a superinflating solution and a quasi-FLRW solution, respectively. The saddle nature of the fixed point $\mathcal{P}$ is consistent with the saddle nature of the fixed point $E$ obtained by Barrow. The fixed point $\mathcal{I}$ obtained by Barrow et.al. in \cite{Barrow:2006xb} is an unstable solution. The addition of an extra dynamical degree of freedom by the Higgs field, and hence two extra dimensions in the phase space, makes the corresponding line of fixed points $\mathcal{L}$ a saddle in the case of the $R^2$-Higgs scenario. 

On the other hand, Jarv et. al. \cite{Jarv:2021qpp} obtain only one finite fixed point $A$ at the minimum of the quartic potential, which corresponds to our fixed point $\mathcal{P}$. Whereas the fixed point is an attractor in \cite{Jarv:2021qpp}, the addition of an extra degree of freedom by the scalaron, and hence two extra dimensions in the phase space, turns the corresponding fixed point into a saddle.

Lastly, we mention again that the fixed points $\mathcal{P}$ and $\mathcal{L}$ can, in a sense, be argued to be asymptotic fixed points, because they lie on the submanifold $x=0$, which lies at the boundary of the physical domain of validity of the phase space. They are, however, \emph{not} asymptotic fixed points in the usual sense of diverging phase space coordinates; hence, no phase space compactification is required for their analysis.

\subsection{Illustrative phase portraits}

\begin{figure}[h]
    \centering
    \includegraphics[width=0.42\linewidth]{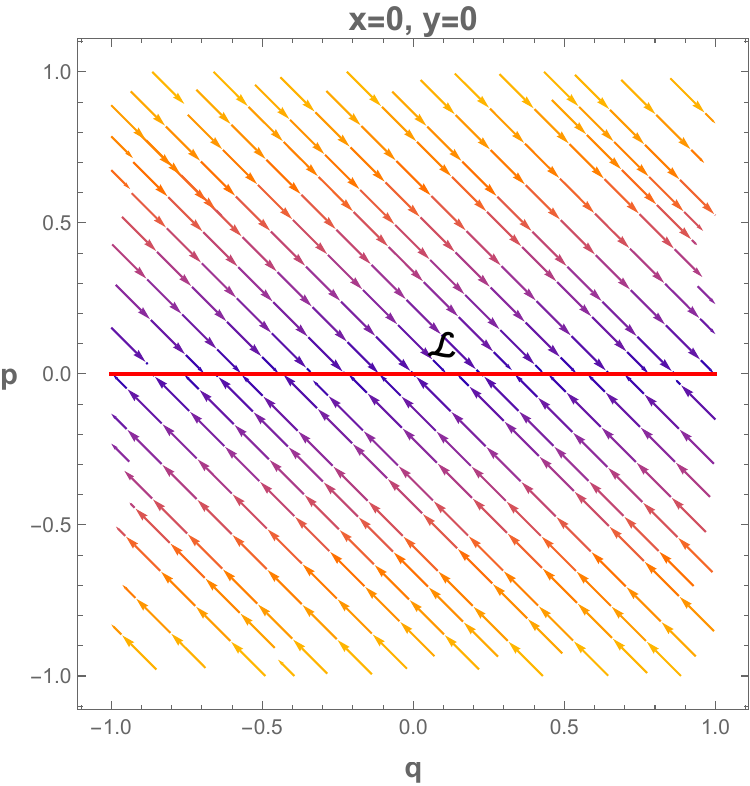}
    \includegraphics[width=0.42\linewidth]{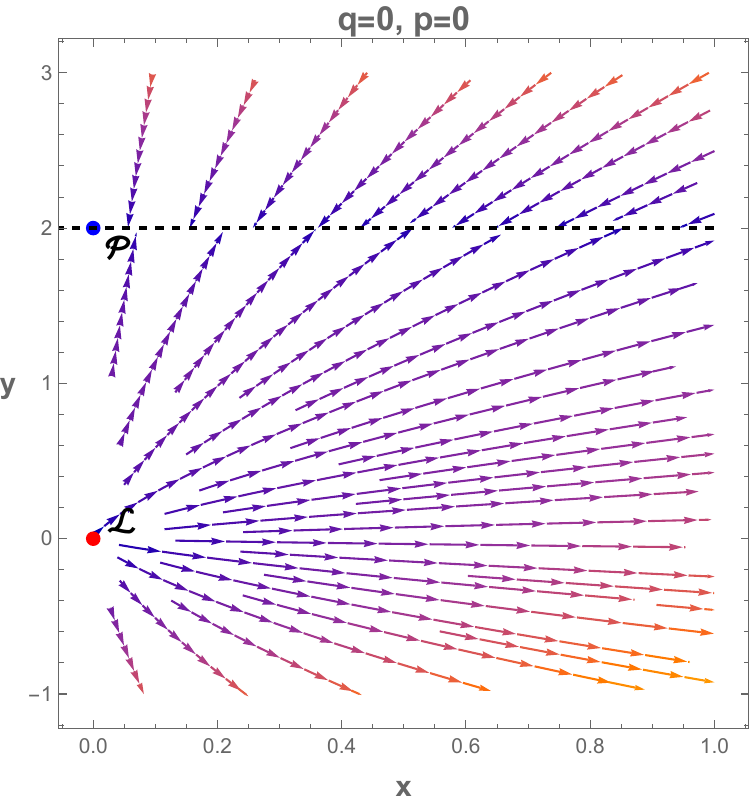}
    \caption{Phase portraits on the 2-dimensional invariant submanifolds $x=0=y$ (the left panel) and $q=0=p$ (the right panel) for the parameter values $\xi=1000,\,\alpha=10^{10}\kappa^2,\,\lambda=0.01$ (taken from \cite{He:2018gyf}). The physical domain of validity of the phase portrait has $x>0$, so the invariant submanifold on the left panel lies at the boundary of the physical phase space. The line of fixed points $\mathcal{L}$ and the isolated finite fixed point $\mathcal{P}$ are indicated wherever possible. There is a center manifold associated with the fixed point $\mathcal{P}$ that is tangent to the line $y=2-\frac{\kappa^2}{6\alpha}x$ (the dashed line in the right-hand panel 
    ) at this point. Note that, since $x=0=y$ and $q=0=p$ are invariant submanifolds, the phase portraits above show original phase trajectories, and not just the projection of original phase trajectories.}
    \label{fig1}
\end{figure}

The complete phase space of the Starobinsky-Higgs inflationary model \eqref{SH_action} is 4-dimensional. In this subsection, we present some illustrative 2-dimensional slices of phase portraits to facilitate an understanding of the phase flow. The first obvious 2-dimensional slices that come to mind are the two 2-dimensional invariant submanifolds $x=0=y$ and $q=0=p$ as was pointed out in subsection \ref{subsec:inv_sub_SH}. The phase portraits on these two 2-dimensional invariant submanifolds are shown in Fig.\ref{fig1}. 

Since they are invariant submanifolds, the respective phase portraits show actual phase trajectories and not just their projections. From Fig.\ref{fig1} (left panel), it is clear that \emph{on} the invariant submanifold $x=0=y$, the line $p=0$ behaves as an attractor, and the trajectories approach it along the direction parallel to the line $p=-q$. This is consistent with the corresponding eigenvalue $-1$ and its corresponding eigenvector; see Table \ref{tab:SH_finite_fp}. 

The line of fixed points $\mathcal{L}$ intersects the invariant submanifold $q=0=p$ at the point $(x=0,y=0)$, which is shown as the red dot in Fig.\ref{fig1} (right panel). Since $q=p=0$ is an actual invariant submanifold, the portrait on the right panel of Fig.\ref{fig1} shows actual trajectories. The existence of a heteroclinic trajectory from $\mathcal{L}$ to $\mathcal{P}$ can be clearly seen from the right panel. However, this may not be the only heteroclinic trajectory connecting $\mathcal{L}$ to $\mathcal{P}$, since $\mathcal{L}$ is actually a line of fixed points. As can be seen from Table \ref{tab:SH_finite_fp}, the fixed point $\mathcal{P}$ is non-hyperbolic and has an associated center manifold which is tangent to the line $y=2-\frac{\kappa^2}{6\alpha}x$ at $\mathcal{P}$. The latter line is indicated in Fig.\ref{fig1}, right panel.

\begin{figure}[H]
    \centering
    \includegraphics[width=0.32\linewidth]{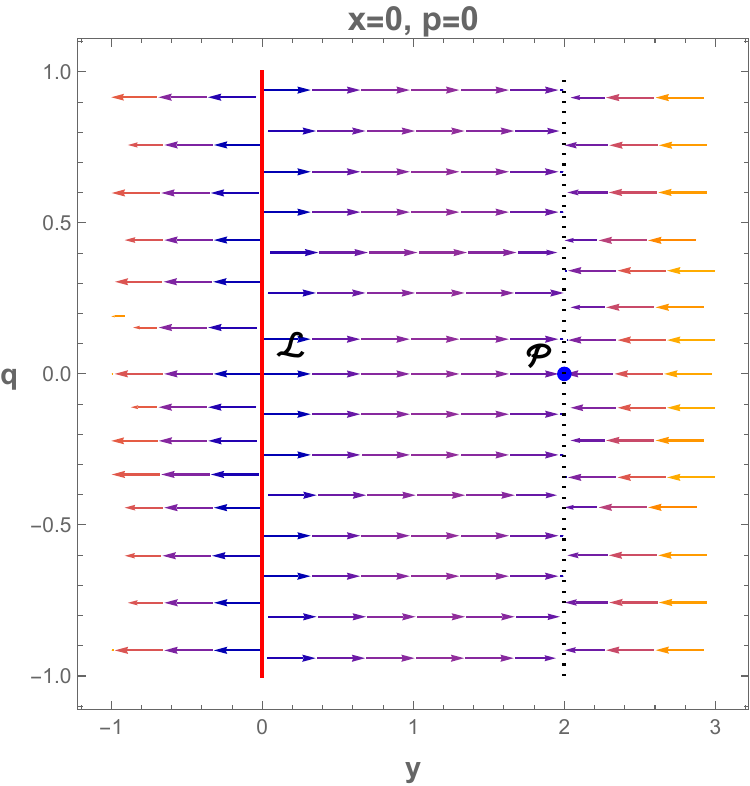}
    \includegraphics[width=0.32\linewidth]{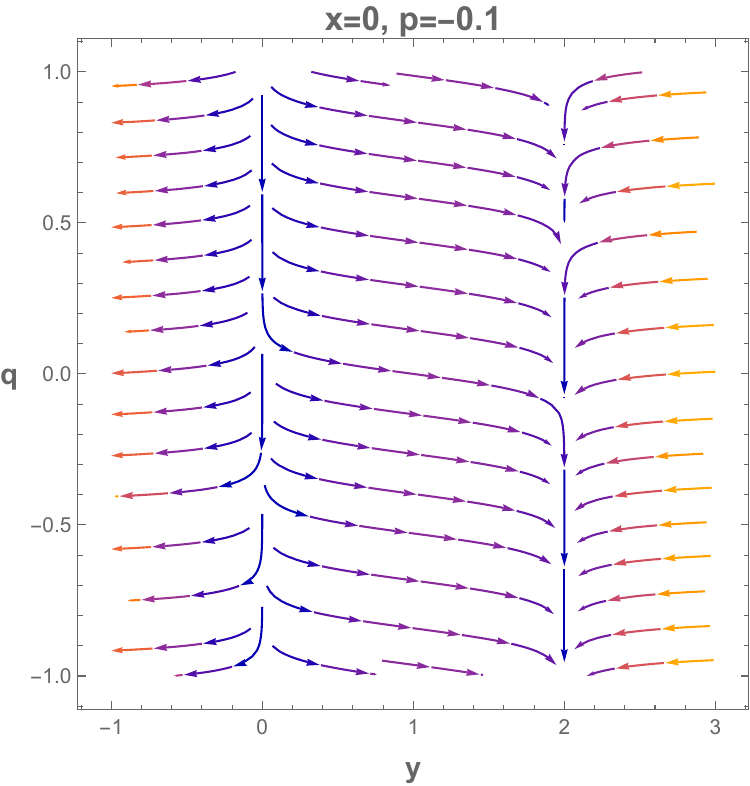}
    \includegraphics[width=0.32\linewidth]{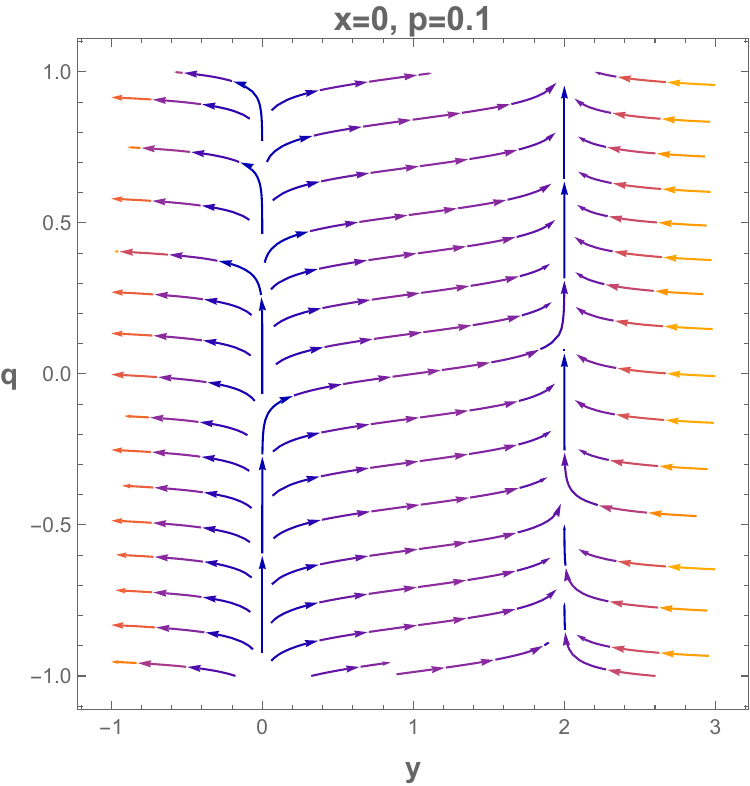}
    \caption{The 2-dimensional projected phase portrait on the slices $x=0,\,p=0$ (left panel), $x=0,\,p=-0.1$ (middle panel) and $x=0,\,p=0.1$ (right panel) for the parameter values $\xi=1000,\,\alpha=10^{10}\kappa^2,\,\lambda=0.01$ (taken from \cite{He:2018gyf}). The left panel shows the line of fixed points $\mathcal{L}$ and the isolated finite fixed point $\mathcal{P}$, both of which lie on the slice $x=0=p$. Although it looks like from that panel that the line $x=q=p=0,\,y=2$ is another line of fixed points, this is \emph{not} actually the case; this is a prime example that projected phase portraits are purely illustrative in nature and does not represent actual phase trajectories (\emph{unless} the slice on which the projection is taken is actually an invariant submanifold). However, since both $\mathcal{L}$ and $\mathcal{P}$ lie on the slice $x=0=p$, the left panel reaffirms the existence of heteroclinic trajectories connecting them. All the slices shown in this figure lie at the boundary of the physical phase space (which has $x>0$).}
    \label{fig2}
\end{figure}

One can notice that the line of fixed points $\mathcal{L}$ and the isolated fixed point $\mathcal{P}$ both lie on the 2-dimensional slice $x=0=p$ (which is \emph{not} an invariant submanifold). This particular slice is shown in Fig.\ref{fig2} (left panel), where $\mathcal{L}$ and $\mathcal{P}$ are also shown. To further visualize the behavior of the trajectories, we have also shown the projected phase portraits on two nearby slices $x=0,\,p=-0.1$ and $x=0,\,p=0.1$ in Fig. \ref{fig2}, middle and right panels, respectively. 

However, since none of the slices shown in Fig.\ref{fig2} are invariant submanifolds, the phase portraits in Fig.\ref{fig2} do \emph{not} show actual phase trajectories, but rather their projections on the respective slices. Therefore, one needs to be careful about interpreting the behavior of the trajectories from this phase portrait; they are purely illustrative projections and \emph{not} actual trajectories. The left panel of Fig.\ref{fig2}, however, reaffirms the existence of a heteroclinic trajectory from $\mathcal{L}$ to $\mathcal{P}$, which was previously seen in the right panel of Fig.\ref{fig1}. All the slices shown in Fig.\ref{fig2} lie at the boundary of the physical phase space (which has $x>0$).

The existence of the heteroclinic trajectory $\mathcal{L}\to\mathcal{P}$ from a quasi-FLRW solution to a superinflationary solution, and the center manifold associated with the super-inflationary solution, is consistent with what was obtained by Barrow et.al. \cite{Barrow:2006xb} for pure quadratic theory (see \cite[Fig.2]{Barrow:2006xb})\footnote{The corresponding trajectory is \cite{Barrow:2006xb} is $\mathcal{I}\to\mathcal{E}$. Because Barrow et.al.\cite{Barrow:2006xb} also had a cosmological constant $\Lambda$ into the picture, they obtained a $\Lambda$-dominated attractor $dS$. The inflationary trajectories passing through $E$ terminated at $dS$. Since we do not have a $\Lambda$ in our picture, there is no additional finite fixed point in our case.}. These are two qualitative features of the pure quadratic inflationary model that survive even after the inclusion of the Higgs field.

 \section{Conclusion}
 \label{summary}

In this work, we developed two unified dynamical system formulations for homogeneous and isotropic cosmology in the general class of $f(R,\phi,X)$ gravity theories. This class includes a wide range of modified gravity and scalar-tensor models, encompassing nonminimal couplings, higher-curvature corrections, and noncanonical kinetic structures. The main innovation of our analysis is the creation of systematic and largely model-independent phase-space frameworks. These frameworks enable the investigation of the global cosmological dynamics for any theory in this class without requiring the development of ad hoc dynamical variables for each specific model.

The first formulation extends the standard Hubble-normalized variables, which are commonly used in $f(R)$ gravity, to the broader case of $f(R,\phi,X)$. While this extension is conceptually straightforward, we found that its applicability depends on two important conditions: the invertibility of the relation that defines the Ricci scalar in terms of the phase-space variables and the non-degeneracy of an auxiliary matrix (\ref{find_R'_X'}) that governs the evolution of $R'$ and $X'$. 
In our analysis, we applied this formulation to a toy model of nonminimal derivative coupling (NMDC) that does not include a scalar potential. We demonstrated that these conditions can indeed be satisfied and that the resulting autonomous system possesses several invariant submanifolds. A significant physical finding from this case is that the qualitative features of the dynamics are independent of the coupling strength between the Ricci scalar $R$ and the scalar field kinetic term $X$. Moreover, we find that the invariant submanifold associated with a frozen scalar field is generally unstable. This latter finding provides a clear dynamical explanation for why, in the absence of a potential, the scalar field does not settle at a constant value: there is no attractor mechanism available to stabilize it. Furthermore, we found that all finite fixed points of this system are non-hyperbolic, which limits the effectiveness of this formulation for a comprehensive stability analysis.

Motivated by these limitations, we introduced a second dynamical system formulation based on an alternative choice of dimensionless variables, in which the Hubble parameter itself plays a central role. One crucial advantage of this approach is that, although it has its own nondegeneracy condition, namely the nondegeneracy of the auxiliary matrix \eqref{find_R'_X'_new}, it is free from any invertibility requirement akin to the first formulation. This significantly broadens the domain of applicability of this formulation as compared to the second formulation. When applied to the mixed $R^2$ Higgs inflationary scenario, this formulation proved robust and physically transparent. In particular, it reproduces the known pure Higgs and pure Starobinsky phase spaces, but these should be understood as degenerate constrained reductions rather than ordinary parameter substitutions. In the Higgs limit, $\alpha=0$ removes the scalaron and the variables $x,y,z$ are fixed by constraint equations in terms of $(q,p)$. In the pure $R^2$ limit, the reduction is obtained by restricting the dynamics to the invariant sector $q=p=0$. These checks confirm the consistency of the formalism on the appropriate lower-dimensional constrained phase spaces.

From a physical perspective, the phase-space analysis of the $R^2$ Higgs model clarifies how the additional scalaron degree of freedom qualitatively alters the dynamics. Fixed points that are attractors in single-field Higgs inflation become saddle points once the scalaron is included, reflecting the enlarged phase space and the interplay between curvature-driven and field-driven inflationary mechanisms. Moreover, we showed that in the large-$\xi$ limit the dynamics effectively reduces to a Starobinsky-like inflationary scenario with a modified scalaron mass, providing a dynamical-systems-based justification of earlier heuristic arguments. The existence of heteroclinic trajectories connecting quasi-FLRW solutions to superinflationary fixed points further demonstrates that key qualitative features of quadratic gravity persist even after the inclusion of the Higgs sector.

Overall, our results establish the second formulation as a general-purpose tool for qualitative cosmological analysis in $f(R,\phi,X)$ gravity. This framework enables a unified investigation of the global phase-space structure, invariant submanifolds, and asymptotic regimes of a broad class of theories, and should be particularly useful for assessing the physical viability of inflationary and dark-energy models where curvature effects, scalar dynamics, and noncanonical kinetic terms interact in a nontrivial manner.

Finally, it is also worth clarifying the scope of the present construction in relation to other dynamical systems recently considered in the literature. The formalism developed here is designed for metric theories whose gravitational Lagrangian can be written in the single-field form $f(R,\phi,X)$, with $X=\frac12\partial_\mu\phi\partial^\mu\phi$. Therefore, models involving more than one scalar field, such as coupled inflaton--phantom systems~\cite{Perez-Lorenzana:2007gog}, are not directly contained in the present phase-space construction. They would require an enlarged set of variables, for instance $q^I=\kappa\phi^I$ and $p^I=\kappa(\phi^I)'$, together with a field-space kinetic structure or several kinetic invariants $X^{IJ}$. Similarly, trace-free Einstein gravity~\cite{Montesinos:2025zat} is a distinct theory, with a different set of field equations and constraint structure, and is not obtained as a particular choice of the function $f(R,\phi,X)$ in Eq.~(1). These theories are nevertheless related to the present work in the broader sense that they also benefit from Hubble-normalized dynamical variables and autonomous phase-space methods. Extending the present construction to multi-field systems or to trace-free gravitational dynamics would therefore be an interesting direction for future work, but lies outside the scope of the single-field $f(R,\phi,X)$ framework studied here.

\appendix

\section{The $f(R)$ dynamical system}\label{app:f(R)}

Consider a situation where the scalar field freezes to a potential minimum, i.e. has no dynamical content. This implies $X=0$ i.e. $p=0$, and any possible potential can be thought of as contributing to an effective cosmological constant, whose value is equal to that of the potential minimum. In such a case, the Lagrangian $f(R,\phi,X)$ effectively reduces to $f(R)$, with the effective cosmological constant term absorbed into the $f(R)$.

We have mentioned in Section \ref{Sec: DS_1} that the dynamical variables $\Omega,x,y,z,K$ defined in Eq.\eqref{variables} are exactly the same as those appearing in the standard $f(R)$ dynamical system formulation (see, e.g. \cite{MacDevette:2022hts}). Let us show below that the dynamical system \eqref{DS} correctly reduces to the $f(R)$ dynamical system when the scalar field freezes.

For a frozen scalar field, $f(R,\phi,X)$ reduces to just $f(R)$. In this case, from Eq.\eqref{g} one can write
\begin{subequations}\label{g_f(R)}
\begin{eqnarray}
&& g_2=\frac{R'}{R}=\frac{f_R}{Rf_{RR}}\frac{f_R'}{f_R}=\Gamma\,z\,,
\\
&& g_1 = \frac{f'}{f} = \frac{f_R}{f}R' = \frac{f_R}{Rf_{RR}}\frac{f_R'}{f_R}\frac{Rf_R}{f} = \Gamma\,\frac{y\,z}{x} \,,
\end{eqnarray}
\end{subequations}
where $\Gamma=\frac{f_R}{R f_{RR}}$. The Friedmann constraint becomes simply 
\begin{equation}
    z = -1 + \Omega - K - x + y\,.\label{const_f(R)}
\end{equation}
One can now substitute $g_1,g_2$ from Eq.\eqref{g_f(R)}, as well as $p=0$, back into the system \eqref{DS}, and utilize the constraint \eqref{const_f(R)} to perform some algebraic manipulations, to obtain
\begin{subequations}\label{DS_f(R)}
\begin{eqnarray}
x' &=& \Gamma\,y\,z + x(2K-2y-z+4) \,,\\
y' &=& y\left(\Gamma\,z-2y+2K+4\right)\,, \\
z' &=& 4-2y+3z-3(1+w)\Omega+4K-y\,z-z^2+z\,K\,, \\
K' &=& -2K (y-K-1)\,, \\
\Omega' &=& -\Omega(-1+3w+z+2y-2K)\,.
\end{eqnarray}
\end{subequations}
This is precisely the $f(R)$ dynamical system, as one can compare with Eq.\cite[Eq.(12)]{MacDevette:2022hts}\footnote{To facilitate the comparison, use the following prescription to change from our variables to those used in \cite{MacDevette:2022hts}: $x\to\tilde{y},\,y\to\tilde{v},\,z\to\tilde{x},\,\Omega\to\tilde{\Omega}\,K\to\tilde{K}$.}

\section{The large $\xi$ limit of the $R^2$-Higgs inflation}\label{app:SH}

Consider the Jordan frame action of the $R^2$-Higgs inflationary model \cite{He:2018gyf}
\begin{equation}
    \mathcal{S} = \int d^4 x \sqrt{-g} \left[ \frac{1}{2\kappa^2}(R + \alpha R^2) + \frac{1}{2}\xi\phi^2 R - \frac{1}{2}\partial_{\alpha}\phi \partial^{\alpha}\phi - \frac{\lambda}{4}\phi^4 \right]\,.
\end{equation}
Writing out the action in the FLRW minisuperspace,
\begin{equation}\label{SH_FLRW}
    \mathcal{S} = \int d^3 x \int dt\,\, a^3 \left[ \frac{1}{2\kappa^2}(R + \alpha R^2) + \frac{1}{2}\xi\phi^2 R + \frac{1}{2}\dot{\phi}^2 - \frac{\lambda}{4}\phi^4 \right]\,,
\end{equation}
let us concentrate on the nonlinear coupling term $\frac{1}{2}\xi\phi^2 R$. One can then perform the following steps
\begin{eqnarray}
    \int dt\,\, a^3 \cdot \frac{1}{2}\xi\phi^2 R &=& \int dt\,\,3\xi\,\phi^2 \left(a^2\ddot{a} + a\dot{a}^2\right) \nonumber\\
    &=& \int dt\,\, 3\xi\,\left[ \frac{d}{dt}(\phi^2 a^2 \dot{a}) - 2\phi\dot{\phi}a^2 \dot{a} - \phi^2 a \dot{a}^2 \right] \nonumber\\
    &=& \int dt\,\,  3\xi\left[ \frac{d}{dt}(\phi^2 a^2 \dot{a}) - \frac{2}{3}\frac{d}{dt}\left(\phi\dot{\phi}a^3\right) - \phi^2 a \dot{a}^2 + \frac{2}{3}a^3(\phi\ddot{\phi} + \dot{\phi}^2)\right] \nonumber\\
    &=& \int dt\,\, a^3 \xi \left[ 2(\dot{\phi}^2 + \phi\ddot{\phi}) - 3H^2 \phi^2 \right]\,,
\end{eqnarray}
where in the last step we have omitted the boundary terms. The FLRW minisuperspace action \eqref{SH_FLRW} then becomes
\begin{equation}
    \mathcal{S} = \int d^3 x \int dt\,\, a^3 \left[ \frac{1}{2\kappa^2}(R + \alpha R^2) + \left(2\xi+\frac{1}{2}\right)\dot{\phi}^2 + 2\xi\phi\ddot{\phi} - 3\xi H^2\phi^2 - \frac{\lambda}{4}\phi^4 \right]\,.
\end{equation}
The above action makes it clear that in the limit of large $\xi$, the original kinetic term $\frac{1}{2}\dot{\phi}^2$ of the Higgs field can be ignored.

If we now ignore the original kinetic term from the FLRW minisuperspace action \eqref{SH_FLRW}, then the resulting action is
\begin{equation}\label{SH_FLRW_large_xi}
   \mathcal{S}_{\text{large $\xi$}} \approx \int d^4 x \sqrt{-g} \left[ \frac{1}{2\kappa^2}(R + \alpha R^2) + \frac{1}{2}\xi\phi^2 R - \frac{\lambda}{4}\phi^4 \right]\,.
\end{equation}
Varying the above action with respect to $\phi$, we get the constraint $\phi^2=\frac{\xi}{\lambda}R$. Implementing this back into the action \eqref{SH_FLRW_large_xi}, we get
\begin{equation}
    \mathcal{S}_{\text{large $\xi$}} \approx \int d^4 x \sqrt{-g}\,\, \frac{1}{2\kappa^2}\left[R + \left(\alpha + \frac{\kappa^2\xi^2}{2\lambda}\right)R^2\right] \,.
\end{equation}
This shows neatly that in the limit of large nonminimal coupling, the $R^2$-Higgs model is effectively just another Starobinsky inflationary scenario with a modified scalaron mass.

\bibliography{Refs}
\bibliographystyle{unsrt}

\end{document}